\documentclass[rmp,aps, twocolumn,showpacs,floatfix,longbibliography]{revtex4-1}
 
\usepackage{amssymb}
\usepackage{natbib}
\usepackage{lipsum}
\usepackage{amsmath}
\usepackage{amsfonts}
\usepackage{graphicx}
\usepackage{mathrsfs}
\usepackage{dcolumn}
\usepackage{bm}
\usepackage{mathbbol}
\usepackage{color}
\usepackage{graphics}
\usepackage{epsfig}
\usepackage{soul,xcolor}
\usepackage[normalem]{ulem}
\def\bb#1{\bm{(}#1\bm{)}}
\def\Tr{\mbox{Tr}}

\begin{document}
\title{{\it Colloquium:}  Statistical Mechanics and Thermodynamics at
Strong Coupling: Quantum and Classical}

\author{Peter Talkner}
\email{peter.talkner@physik.uni-augsburg.de} 
\affiliation{Institute of Physics, University of Augsburg,
  Universit\"atsstra{\ss}e 1, D-86135 Augsburg, Germany\\
Center for Theoretical Physics of Complex Systems, Institute for Basic Science (IBS), Daejeon, 34113, Republic of Korea }
\author{ Peter H\"anggi}
\email{hanggi@physik.uni-augsburg.de}
\affiliation{Institute of Physics, University of Augsburg,
  Universit\"atsstra{\ss}e 1, D-86135 Augsburg, Germany\\ 
Nanosystems Initiative Munich, Schellingstr. 4, D-80799 M\"unchen, Germany\\
Department of Physics, National University of Singapore, Singapore 117546}
\date{\today}

\begin{abstract}
The statistical mechanical description of small systems staying in thermal equilibrium with an environment can be achieved by means of the Hamiltonian of mean force. In  contrast to the reduced density matrix of an open quantum system, or the reduced phase-space probability density function of a classical open system,  the Hamiltonian of mean force not only characterizes the reduced state but also contains full information about the thermodynamics of the considered open system. The resulting thermodynamic potentials all assume the form as the difference of the potentials for the total system and the bare environment in the absence of the system. In contrast to work as a mechanical notion, one faces several  problems with the definition of heat which turns out to be largely ambiguous in  the case of strong coupling between system and environment. The  general theory of the thermodynamics of open systems, in particular, in view of strong coupling, is reviewed and illustrated  with several examples. The vagueness of heat is discussed in the context of the ambiguities in the definitions of a fluctuating internal energy and other fluctuating  thermodynamic potentials.              
\end{abstract}

\maketitle

\tableofcontents

\section{Introduction}
\label{RMP_1}
Thermodynamics was mainly developed in the 19th century~\cite{vonLaue50Book}
as a phenomenological theory characterizing equilibrium states of macroscopic bodies and their transformations. In spite of the tremendously large number of microscopic degrees of freedom of a macroscopic system, the number of variables characterizing a thermodynamic equilibrium state is extremely small. For a homogeneous system consisting of a single chemical species, the energy, the mole-number and the volume taken by the system uniquely specify the equilibrium state \cite{Callen85Book}. These variables determine the entropy of the system, which is an extensive function, i.e. a homogeneous function of degree 1,  of the variables, provided that gravitation if present at all, can be treated as an external field, but does not play a role as an internal interaction. Self-gravitating and other systems with long-range interactions\footnote{\label{fn_Coulomb}Even though the bare Coulomb potential of an electrical charge decays in the same way as the gravitational potential, the presence of opposite charges leads to screening such that the thermodynamics of (on the whole neutral) systems consisting of charged constituents does not differ from systems with short-range interactions~\cite{Lieb72AM}.} deviate strongly in their behavior from ``normal'' systems~\cite{Kubo65Book}.
For example, the entropy of a self-gravitating system is no longer  a concave  function of the energy everywhere~\cite{Thirring03PRL}.           

Statistical mechanics, on the other hand, provides  a quantum-mechanical~\footnote{\label{fn_classical}For simplicity, in the Introduction we refer mostly to classical systems. Mutatis mutandis, all statements made here also apply to quantum systems. A distinction between classical and quantum systems is made in later sections, where it becomes necessary to do so.} foundation of thermodynamics and yields methods to determine the thermodynamic potentials, such as the entropy, internal, and free energy  for specific systems. In principle, these potentials depend on the kind of contact between the considered system and its environment, whether the  respective system is thermally isolated, or allows an exchange of heat or  particles.  
Technically speaking, ensemble equivalence is tantamount to the concavity of the  entropy as a  function of the energy. In particular this property leads to a positive specific heat. Moreover, the transition between different ensembles, say from the microcanonical to the canonical ensemble, is one to one, given by a Laplace transformation of the density of states. In the thermodynamic limit, this Laplace transformation can be calculated using a steepest descent approximation~\cite{Fowler36Book,Ellis85Book,TouchettePR09,TouchetteEPL11} relating the internal energy to the free energy in terms of a Legendre transformation.  

For isolated, finite systems 
the familiar relations valid in the thermodynamic limit of systems with short-range interactions no longer need to be satisfied. Prevailing finite size corrections, depending on the form of the system such as on the presence of corners, surfaces, and their curvatures, enter the density of states and give rise to deviations from       
the extensivity of the thermodynamical potentials \cite{Baltes76Book}. 
For a discussion of the proper thermodynamic description of isolated, i.e. microcanonical, systems with a finite and possibly small number of degrees of freedom,  we refer to the literature~\cite{Schluter48ZN,Lustig94JCP,Dunkel06PhysA,Hilbert14PRE,Campisi15PRE,Hanggi16PTRS}.

A small, weakly interacting  part of a large system in a microcanonical state approaches a Maxwell-Boltzmann distribution in the thermodynamic limit of the total system~\cite{Khinchin49Book}.     
Any weakly interacting part of a finite normal system staying in a canonical equilibrium state at some temperature is described by a Maxwell-Boltzmann distribution at the temperature of the total system.  
It is important to note that under these conditions the only consequence of the presence of the environment is to maintain the equilibrium of the considered small system at a specified temperature. Accordingly, the phase-space distribution of the small system is independent of any  properties  of the environment other than the temperature and gives rise to the standard thermodynamics of a canonical system independently of its size or form.\footnote{When particles can be exchanged with the environment, a chemical potential is also needed for each particle sort  to characterize the environment.}  
This universality is lost as soon as one considers the total system as a finite microcanonical system at a specific energy even if the interaction between the proper system and its environment is arbitrarily weak~\cite{Campisi07PL}. 

In this Colloquium we restrict ourselves to open systems that are possibly strongly interacting parts of larger systems  with which they stay in canonical equilibrium at a given  temperature. Based on the ergodic hypothesis,~\cite{Lebowitz73PT} this state of the total system can be realized in  different ways, either as an ensemble of microcanonical systems at different energies with an exponential distribution, or as a single open system weakly coupling to a superbath at the prescribed temperature.  Even though these two scenarios yield  identical phase-space distributions of the open system they are not completely equivalent to each other because a cyclic change of a system parameter within a finite time leads to a change of the ensemble representing the total system at large times while, in the second case, the total system then returns to its initial state due to the presence of the superbath~\cite{Talkner16bPRE}.

For total systems, which are prepared in either way in a canonical state, the reduced state of the open system may still be written in the form of a Boltzmann distribution at the temperature of the total system with a {\it Hamiltonian of mean force}~\cite{Onsager33,Kirkwood35JCP,Hanggi90RMP,Jarzynski04JSM,Campisi09PRL} replacing the bare Hamiltonian of the system in the limit of weak coupling. In contrast to the bare system Hamiltonian, the Hamiltonian of mean force depends on temperature and also  other parameters determining the microscopic behavior of the environment and its interaction with the  open system. While the Hamiltonian of mean force completely specifies the reduced state of the open system, the knowledge of this reduced state in the form of a phase-space probability density function is not sufficient to specify the Hamiltonian of mean force. Therefore, the Hamiltonian of mean force cannot be obtained from a purely system-intrinsic point of view, as discussed in more detail in Sec.~\ref{Hmf}. It was argued~\cite{Aurell18PRE} that with the embedding techniques of nonlinear dynamics~\cite{Kantz97Book,Badii94RMP} the  Hamiltonian of the total system can be inferred from observed trajectories of the open system. Given the enormous number of environmental degrees of freedom together with the large amount of data that are required to estimate an unknown Hamiltonian of a system with a few degrees of freedom finding the Hamiltonian of the total system in this way presents in practice an impossible task. For quantum systems even a formal procedure corresponding to the embedding technique of classical dynamical systems is missing.

The statistical mechanics of an open system in equilibrium  can be specified  in the standard way by the Gibbs distribution with  the Hamiltonian of mean force, thereby replacing the bare system Hamiltonian. The according partition function of the open system is given by the ratio of the partition functions of the total system and the bare environment~\cite{Hanggi06APPB}. Consequently, all thermodynamic functions of an open system and their derivatives are specified as differences of the respective quantities of the total system and the bare environment. Further, the resulting thermodynamic potentials are {\it thermodynamically consistent}~\cite{Seifert16PRL,Talkner16bPRE} in a sense that is defined later. However, a possible temperature dependence of the Hamiltonian of mean force  leads to additional contributions to the statistical mechanical expressions of the internal energy and the thermodynamic entropy, which, in general, does not assume the form of a Shannon or von-Neumann entropy \cite{vonNeumann55Book,Wehrl78RMP}. As a consequence the thermodynamic entropy of an open system need {\it not} be a functional of the reduced state of the open system \cite{Seifert16PRL,Talkner16bPRE}; see  Eq.~(\ref{KLS}). 

While fluctuations of macroscopic quantities are generally extremely small in macroscopic systems at thermal equilibrium,\footnote{\label{fn_orderpar}Order-parameter fluctuations at second order phase transitions provide a known exception to this rule.} one may expect that they cannot be neglected in microscopic or mesoscopic systems  and hence become an important issue in this context. Because of their time dependence they convey dynamic information about systems in equilibrium.      
In classical systems one may often identify a set of variables that undergo a Markovian time evolution~\cite{Risken89Book,Gardiner85Book,vanKampen07Book,Stratonovich63Book,Hanggi82PR}. For the description of the energetics of such systems, fluctuating heat, work, and internal energy were introduced as ``Stochastic Energetics'' \cite{Sekimoto10Book}. With the definition of  stochastic entropy \cite{Seifert12RPP} and of further fluctuating thermodynamic potentials \cite{Jarzynski17PRX,Seifert16PRL} a ``Stochastic Thermodynamics'' was established recently. Stochastic energetics already suffers from the  existence of many random functions, for which the thermal equilibrium averages agree with the correct internal energy of the considered open system. The same flaw also adheres to stochastic thermodynamics because thermodynamic consistency is not sufficient to remove this nonuniqueness. Other restrictions on the hypothetical fluctuating thermodynamic potentials are not known~\cite{Talkner16bPRE} for systems other than those weakly coupling to their environments~\cite{Talkner09JSM}.      
For quantum systems the work performed on an open system in an individual run of a force protocol can, in principal, be obtained as a fluctuating quantity by means of two projective energy measurements. But even if the experimental techniques to perform projective measurements, such as nondemolition measurements~\cite{Braginsky80Sci,Yang19arX}, can be substantially improved, the fact that the work is given by the difference of two often extremely large numbers is seriously limiting the practical accessibility of fluctuating work by means of the two point projective energy measurement scheme (TPPEMS).
 
For classical systems, the energy difference may be expressed as an integral of the power supplied to the system, and the power can be determined from an observation of the proper system alone~\cite{Liphardt01SC}.    
But even if the work supplied to a classical open system is known, an unambiguous identification of  heat, i.e., of the energy that is exchanged within the same process between the system and its environment, is possible only in the weak coupling limit, in which the system-environment interaction is all but neglected. This ambiguity of fluctuating heat is also inherent in the notion of fluctuating energy as the sum of fluctuating work and heat, according to a corresponding formulation of the first law. 

The characterization of heat in quantum systems may in principle be based on a TPPEMS of a conveniently defined energy operator of the heat bath, which, once added to the open system internal energy operator, yields the total system Hamiltonian. In processes with a finite interaction time between system and environment, the total transferred heat can be determined by measurements of the bath Hamiltonian before the interaction with the system sets in and after it has ended  \cite{Goold14PRE}. 
In the case of weak coupling, this environmental energy operator coincides with the bare bath Hamiltonian up to a negligibly small contribution of the system-bath interaction. In all other cases the nonuniqueness of the open system internal energy operator also renders the heat-bath energy operator ambiguous. 
But also with an arbitrary specification of the bath energy operator, a joint measurement of this operator and the total Hamiltonian {\it cannot} be achieved because of their noncommutativity. Hence, it is not possible to simultaneously specify for a quantum process  work and heat, not even their averages~\cite{Hanggi15NP,Castelvecchi17N}.     

\section{Thermodynamics of large normal systems}\label{RMP_2}
We first summarize the thermodynamics and statistical mechanics of ``normal'' systems~\cite{Kubo65Book} with the goal of recollecting the notions relevant for our main discussion and  introducing the notation. 

To start with, we refer to a system  consisting of a macroscopically large number of microscopic objects like atoms or molecules as  {\it normal } if in the  quantum case the logarithm of the number of states, and in the classical case the logarithm of phase-space volume below a given energy, are homogeneous convex functions of the extensive variables~\cite{Kubo65Book}.   
To satisfy this requirement for a system of classical particles experiencing pairwise interactions in $d$ spatial dimensions,  the interaction potential must be repulsive at short distances  and decay with the distance $r$ faster than $r^{-d}$~\cite{Ruelle69Book,Lieb72AM}. Further, we  assume that the dynamics of an autonomous, isolated normal system approaches, after a sufficiently long time, a unique equilibrium state that is independent of  the initial state of the system other than its energy. For classical systems this is guaranteed by ergodicity~\cite{Khinchin49Book}, for quantum systems the problem of thermalization as such has long been recognized~\cite{vonNeumann29ZP} but is still under active scrutiny~\cite{Polkovnikov11RMP,Goold16JPhysA,Srednicki94PRE,Deutsch91PRA,Rigol12PRL,Merali17N}. In this Colloquium we shall  not extend this discussion.

\subsection{Isolated systems}
\label{isolated}
The dynamics of an isolated system is governed by a Hamiltonian which is the Hamilton function in the case of a classical system and the Hamilton operator for quantum systems. We assume that the gauge of the Hamiltonian is chosen in a way such that it yields the energy of the system even if the parameters $\lambda$ specifying the Hamiltonian depend on time~\cite{Goldstein02Book}. At any fixed set of parameter values the system is supposed to approach an equilibrium state that is completely specified by the energy of the system and hence given by~\cite{Bopp53ZNA,Munster54ZP}
\begin{equation}
\rho = \omega^{-1}(E,\lambda) \delta \bb{ E-H(\lambda) }\:.
\label{rmic}
\end{equation}
For a quantum system $\rho$ presents a density matrix, i.e., a positive operator on the system's Hilbert space with unite trace, and $\delta$ denotes the Dirac delta function. The normalization of the density matrix is guaranteed by the inverse of the density of states, which is given by
\begin{equation}
\begin{split}
\omega(E, \lambda) &= \Tr \delta\bb {E-H(\lambda) }\\
&= \sum_n d_n \delta\bb {E-E_n(\lambda) }\:,
\end{split}
\label{omegaqm}
\end{equation}
where $E_n(\lambda)$ are the eigenvalues of the Hamiltonian $H(\lambda)$ and $d_n$ are the corresponding degrees of degeneracy.\footnote{\label{fn_mcspectrum}The density of states is not defined for energies from the continuous part of the spectrum. Using the spectral representation of the Hamiltonian one formally obtains  with $\delta\bb {E-H(\lambda) } = \int \delta(E-E') dP(E')$  the expression $\omega(E,\lambda) = \int \delta(E-E') \Tr dP(E')$, yielding for an energy belonging to the point spectrum with $\Tr dP(E_n) = d_n$ the result given in Eq. (\ref{omegaqm}). Here $P(E)$ denotes the projection operator onto the subspace spanned by all eigenfunctions of $H$ with energies up to $E$.   For an energy from the continuous spectrum the trace expression is given by the squared norm of the corresponding eigenfunction, which diverges.} Because of the discreteness of the energy spectrum a regularized form of the delta function entering the density matrix must be considered, such as a narrow Gaussian function $\delta_\epsilon(x) =(2 \pi)^{-1/2} \epsilon \exp \bb { - x^2/(2 \epsilon) }$~\cite{Talkner08PRE}.

For a classical system $\rho(x)$ presents the probability density function (pdf). It takes the  same form as Eq.~(\ref{rmic}), with $H(\lambda) = H(\mathbf{x},\lambda)$ denoting the Hamilton function; the density of states then becomes
\begin{equation}
\omega(E,\lambda) = \int d \mathbf{x} \delta \bb {E-H(\mathbf{x},\lambda) }\:,
\label{classical}
\end{equation}
with a conveniently defined dimensionless infinitesimal phase-space volume $d\mathbf{x}$, which allows for indistinguishable particles if necessary, such as $d\mathbf{x} = d^{3N}p d^{3N}q/(N! h^{3N})$ with the Planck constant $h$ in the case of $N$ particles in three dimensions, exhibiting no further symmetries.
The thermodynamics of a system in a microcanonical state [Eq.~(\ref{rmic})] is determined by the microcanonical entropy $S(E,\lambda)$, which is given 
by \cite{Gibbs02Book,Hertz10APLa,Hertz10APLb,Hilbert14PRE}
\begin{equation}
S(E,\lambda)= k_B \ln \Omega(E,\lambda)\:,
\label{Smc}
\end{equation}
where $k_B$ is the Boltzmann constant and $\Omega(E,\lambda)$ specifies the number of states of a quantum system or the phase-space volume of a classical system below the energy $E$ and hence reads 
\begin{equation}
\begin{split}
\Omega(E,\lambda) &= \int_0^E dE'\omega(E',\lambda)\\
&= \Tr \Theta\bb {E-H(\lambda) }\qquad \quad \; \text{quantum}\:,\\
&= \int d\mathbf{x} \Theta\bb {E-H(\mathbf{x},\lambda)}\quad \text{classical}\:.
\end{split}
\label{Oo}
\end{equation}
Here $\Theta(x)$ denotes the Heaviside function that vanishes for negative arguments $x$ and yields unity for positive ones.  
The total differential of the entropy is given by
\begin{equation}
d S = \frac{1}{T} dE + \sum_n \frac{a_n}{T} d\lambda_n\:,
\label{dS}
\end{equation}
where $T= (\partial S/\partial E)^{-1}_\lambda$ designates the microcanonical temperature and $a_i = T (\partial S/\partial \lambda_i)_{E, \lambda_{j\neq i}}$ specifies the response coefficient for a variation of the parameter $\lambda_i$.  Most notably, the thermodynamic expression for $a_i$ coincides with its statistical mechanical definition given by the microcanonical average $a_i = \langle \partial H(\lambda)/\partial \lambda_i \rangle$~\cite{Dunkel14NP,Hilbert14PRE}. This consistency of statistical mechanics and thermodynamics is guaranteed only for an entropy that depends on the total phase-space volume, as in Eqs. (\ref{Smc}) and (\ref{Oo}).\footnote{The logarithmic dependence of the entropy on the phase-space volume is a consequence of the additivity of the entropy for noninteracting systems.} It agrees~\footnote{Up to corrections of the order of $o(N)$, with $N$ denoting the number of microscopic degrees of freedom. } for normal systems with the more familiar entropy definition in terms of the density of states, reading
\begin{equation}
S_B(E,\lambda) = k_B \ln \left [ \omega(E,\lambda) \epsilon_B \right ]\:,
\label{SBmc}
\end{equation} 
where $\epsilon_B$ is an energy scale that must not depend on the values of $E$ and $\lambda$.  
We note that for quantum systems the phase-space volume entropy~(\ref{Smc}) is a piecewise constant function of the energy and hence must be smoothed to yield a well-defined temperature and, more generally, to serve as a thermodynamic quantity. The necessary interpolation of the entropy for energies that are different from the eigenvalues of the system Hamiltonian, though, introduces a certain ambiguity.  
\subsection{Small subsystem of a large closed system}\label{bigmicro}    
The energy of a subsystem fluctuates even if the total system has a fixed energy. The energy fluctuations of an open system follow a Boltzmann distribution  provided that the interaction between the considered part and the total system is weak in the sense that the interaction energy is much smaller than the average energy of the subsystem. Additionally, the total system must be much  larger than the subsystem\footnote{\label{fn_Khinchin} When a large  system is subdivided into two equally large parts, the energy of each part is approximately Gaussian distributed~\cite{Khinchin49Book}.}. The temperature of the Boltzmann distribution of a small system is determined by the microcanonical Boltzmann temperature\footnote{\label{fn_Boltzmann}The Boltzmann temperature $T_B(E)$ follows from the Boltzmann entropy~(\ref{SBmc}) as $T_B(E) 
=(\partial S_B(E,\lambda)/\partial E)^{-1}_\lambda$. }   calculated at the average value of the energy of the large part that provides the environment of the considered open system \cite{Hanggi16PTRS}. In the thermodynamic limit, this temperature agrees with the microcanonical temperature of the total system provided that the latter is normal; i.e., $T=T_B$.

Under these conditions, the state of an open system is given by 
\begin{equation}
\rho(\beta,\lambda) = Z^{-1}(\beta,\lambda) e^{-\beta H_S(\lambda)}\:,
\label{rcan}
\end{equation}
where $\beta= (k_B T)^{-1}$ is the inverse temperature and $H_S(\lambda)$ is the Hamiltonian operator or function of the isolated quantum or classical subsystem, respectively. For a quantum system $\rho(\beta,\lambda)$ denotes the density matrix and for a classical system the function $\rho(\mathbf{x},\beta,\lambda)$ specifies the phase-space probability density function at the phase-space point $\mathbf{x}$. The dimensionless partition function $Z(\beta,\lambda)$ serves for normalization and hence reads\footnote{For the partition function to exist, the Gibbsian operator $e^{-\beta H_S(\lambda)}$ must be an  element of the trace class~\cite{Schatten50Book}, which is tantamount to the requirement that the Gibbsian operator and consequently also the Hamiltonian have a pure point spectrum but does not contain an absolute or singular continuous part. Moreover, the Hamiltonian must be bounded from below. For classical systems the potential energy must be sufficiently confining  to prevent the system from escaping to infinity and also bounded from below.}
\begin{align}
\label{Zqm}
Z(\beta,\lambda) &= \Tr e^{-\beta H_S(\lambda)} \quad &\text{quantum}\:,\\
&=\int d\mathbf{x} e^{-\beta H_S(\mathbf{x},\lambda)} &\text{classical}\:.
\label{Zcl}
\end{align} 

With the knowledge of the partition function the connection between  statistical mechanics and thermodynamics is established by 
\begin{equation}
F(\beta,\lambda) = - \beta^{-1} \ln Z(\beta,\lambda)\:,
\label{FZ}
\end{equation}
which defines the free energy $F(\beta,\lambda)$. From this point on, other thermodynamic potentials such as the internal energy $U(\beta,\lambda)$ and the entropy $S(\beta,\lambda)$  can be obtained in terms of the text book relations~\cite{Callen85Book}
\begin{align}
U(\beta,\lambda) &= \left .\frac{\partial \bb { \beta F(\beta,\lambda) }}{\partial \beta}\right |_\lambda\:, 
\label{U}\\
S(\beta,\lambda) &= \left . k_B \beta^2 \frac{\partial F(\beta,\lambda)}{\partial \beta} \right|_\lambda\:,
\label{S}
\end{align}
which are connected by 
\begin{equation}
F = U - S T\:.
\label{FUS}
\end{equation}
The joint validity of the three relations (\ref{U})--(\ref{FUS}) constitutes the {\it thermodynamic consistency} of the thermodynamic potentials   $F(\beta,\lambda),\; U(\beta,\lambda)$ and $S(\beta,\lambda)$~\cite{Seifert16PRL,Talkner16bPRE}. Any two of the three relations imply the remaining  one. 

In the case of a {\it weakly coupled open system} in thermal equilibrium, the internal energy and the entropy agree with the standard statistical mechanical expressions  for systems in canonical equilibrium, i.e.
\begin{align}
U(\beta,\lambda) &= \Tr H_S(\lambda) \rho_S(\beta,\lambda)\:, 
\label{Ucan}\\
S(\beta,\lambda) &= - k_B\Tr \rho(\beta,\lambda) \ln \rho(\beta,\lambda)\:,
\label{Scan}
\end{align}
expressing the internal energy $U(\beta,\lambda)$ as the average value of the bare system Hamiltonian $H_S$ with respect to the canonical equilibrium state $\rho(\beta,\lambda)$ specified in Eq. (\ref{rcan}) and the entropy $S(\beta,\lambda)$ as the von Neumann entropy for quantum systems or the Gibbs-Shannon entropy for classical systems. For classical systems, in  Eqs. (\ref{Ucan} and \ref{Scan}) the trace is to be replaced by the phase-space integral ($\Tr \rightarrow \int d\mathbf{x}$) and the density matrix by the according phase-space pdf.  

For the discussion of a small system weakly coupling to a finite bath see~\citet{Campisi09PRE}.
\section{Subsystem of a total system at equilibrium in a canonical state}\label{RMP_3}
Any canonical equilibrium state can in principle be realized in two physically different ways. As described   
in Sec.~\ref{bigmicro}, the canonical state may result from the weak contact with another, much larger, system. The time average of the ever changing state of the open system is then given by the canonical state. The other, more formal, way to consider a canonical state, is to interpret it as an ensemble of microcanonical states  the energies of which follow a Boltzmann distribution at the given inverse Boltzmann temperature $\beta$. 
While in the first interpretation one considers a single open system, the second scenario consists of many closed systems.

\subsection{Hamiltonian of mean force}\label{Hmf}
The Hamiltonian of mean force \cite{Campisi09PRL} is a fundamental concept in the study of an open system that stays together with its environment in a canonical equilibrium state. It generalizes the notion of the potential of mean force~\cite{Kirkwood35JCP,Roux95CPC,Hanggi90RMP,Onsager33} and, in contrast to the latter, it is not restricted to classical situations and can also be assigned to an open quantum system. We first present its definition for quantum systems and later specialize to classical ones.

The starting point is the Hamiltonian $H_{\text{tot}}$ of the total system which is composed of contributions describing the bare system and the bare environment, $H_S(\lambda)$ and $H_B$, respectively, and an interaction-term $H_{SB}$. As before, the system Hamiltonian is assumed to depend on a set $\lambda$ of controllable parameters that must have no influence on the bare environmental Hamiltonian or the interaction.\footnote{This restriction is not  necessary here but becomes essential for the definition of work done on an open system by a variation of the parameter $\lambda$; see Sec.~\ref{open}. Other globally acting parameters specifying the Hamiltonian of the total system are generally  suppressed so as not to overburden the notation.} The total Hamiltonian is therefore given by
\begin{equation}
H_{\text{tot}}(\lambda) = H_S(\lambda) +H_B + H_{SB}\:.
\label{Htot}
\end{equation}
While the canonical thermal equilibrium state of the total system   
follows with eq. (\ref{rcan}) as
\begin{equation}
\rho_{\text{tot}}(\beta,\lambda) = Z^{-1}_{\text{tot}}(\beta,\lambda) e^{- \beta H_{\text{tot}}(\lambda)}\:,
\label{rtot}
\end{equation}
the state of the open system is determined by the reduced density matrix $\rho_S(\beta,\lambda)$ and hence becomes
\begin{equation}
\rho_S(\beta,\lambda) = Z^{-1}_{\text{tot}}(\beta,\lambda) \Tr_B e^{-\beta H_{\text{tot}}(\lambda)}\:,
\label{rS}
\end{equation}
where $\Tr_B$ denotes the partial trace over the environmental Hilbert space.
The reduced density matrix is proportional to the ``renormalized Boltzmann-factor'' $e^{-\beta H^*(\beta,\lambda)}$ with the Hamiltonian of mean force $H^*(\beta,\lambda)$. This renormalized system's Boltzmann-factor results from the Boltzmann-factor of the total system  by a properly normalized partial trace generating an average over all environmental configurations according to their occurrence in thermal equilibrium, given by
\begin{equation}
e^{-\beta H^*(\beta,\lambda)} = Z^{-1}_B(\beta) \Tr_B e^{-\beta H_{\text{tot}}(\lambda)}\:.
\label{eH*}
\end{equation}
The normalization with the partition function of the bare environment, given by
\begin{equation}
Z_B(\beta) = \Tr_B e^{-\beta H_B}\:,
\label{ZB}
\end{equation}
is {\it uniquely} determined by the requirement that for a vanishing system--environment interaction the renormalization yields the  result
\begin{equation}
e^{-\beta H^*(\beta,\lambda)} = e^{-\beta H_S(\lambda)} \quad \text{for}\; H_{SB}=0\:,
\label{H*0}
\end{equation}  
or, equivalently, $H^*(\beta,\lambda) =H_S(\lambda)$ for the vanishing interaction $H_{SB}$. 
Finally, the reduced state of the open system can be expressed in terms of the Hamiltonian of mean force, yielding 
\begin{equation}
\rho_S(\beta,\lambda) = Z^{-1}_S(\beta,\lambda) e^{-\beta H^*(\beta,\lambda)}\:,
\label{rSH*}
\end{equation}
with
\begin{equation}
H^*(\beta,\lambda) = - \beta^{-1} \ln \frac{\Tr_B e^{-\beta H_{\text{tot}}(\lambda)}}{\Tr_B e^{-\beta H_B}}\:.
\label{qH*}
\end{equation}

Comparing Eqs. (\ref{rS}) and (\ref{rSH*}) one finds that the partition function of the open system is given by the ratio of the partition functions of the total system and the bare environment, i.e.,
\begin{equation}
Z_S(\beta,\lambda) = \Tr_S e^{-\beta H^*(\beta,\lambda)} = \frac{Z_{\text{tot}}(\beta,\lambda)}{Z_B(\beta)}\:.
\label{ZSZtotZB}
\end{equation}
where $\Tr_S$ denotes the trace over the Hilbert space of the system. The requirement that the renormalization procedure reproduces the bare system Boltzmann factor for a vanishing system-bath interaction was missing in~\citet{Gelin09PRE}, leading to the erroneous conclusion that the  particular form of the partition function as a ratio is arbitrary.    

Before we discuss the consequences of this particular structure of the partition function $Z_S$ for the thermodynamics of an open system, we  emphasize the following facts:\\
(i) As indicated by the notation, and as made explicit in specific examples discussed later, the Hamiltonian of mean force depends not only on the parameters $\lambda$ entering the bare system Hamiltonian  but, in general, also on the temperature of the total system, and on all other global parameters as well. \\
(ii) Moreover, the structure of the Hamiltonian of mean force depends on the type of environment and its interaction with the open system. Beyond the case of weak coupling, which is considered later, one cannot assume the existence of a generic ``thermal environment'' the details of which were irrelevant. In general, instead, different environments lead to different Hamiltonians of mean force for the same bare system. \\   
(iii) Further, we emphasize that the Hamiltonian of mean force does not follow from  the reduced state of the open system.  From a known, reduced density matrix $\rho_S(\beta,\lambda)$ with the help of eqs. (\ref{rSH*}) and (\ref{ZSZtotZB}) one can determine the sum of the Hamiltonian of mean force and the Helmholtz free energy of the open system as
\begin{equation}
H^*(\beta,\lambda) + F_{\text{tot}}(\beta,\lambda) - F_B(\beta) = - \beta^{-1} \ln \rho_S(\beta,\lambda)\:,
\label{H*F}
\end{equation}   
where we express the logarithms of the partition functions in terms of the respective free energies; see Eq. (\ref{FZ}). If there is no additional knowledge about these free energies, the Hamiltonian of mean force remains undetermined.\\ 
(iv) Finally, we note that the partition function $Z_S(\beta,\lambda)$ defined by Eq. (\ref{ZSZtotZB}) remains finite in the thermodynamic limit of the environment, whereby the number of system degrees of freedom is kept fixed. In this limit the partition functions of the total system as well as that of the environment either diverge or vanish with an increasing number of environmental degrees of freedom \footnote{\label{fn_normal} We assume here that the total system is normal in the sense of \cite{Kubo65Book}  as previously explained.}, yet their ratio remains finite. In the theory of stochastic energetics~\cite{Sekimoto10Book} and also in a recent review on stochastic thermodynamics~\cite{Seifert19ARCMP} coarse-grained, still fluctuating, free energies are introduced without subtraction of the  bath contribution. This does not impact free energy differences in an  isothermal process, but  for nonisothermal processes or processes with globally changing parameters it    leads to extra contributions to the internal energy and the entropy, as discussed later in more detail.      

Equation (\ref{qH*}), which specifies the Hamiltonian operator of mean force for a quantum system, can immediately be adapted to classical systems by replacing the Hamiltonian operators with the respective Hamiltonian functions, and the partial trace over the environment with an integral over the environment's phase space, yielding
\begin{equation}
e^{-\beta H^*(\mathbf{x},\beta,\lambda)}= e^{-\beta H_S(\mathbf{x},\lambda)} \int d\mathbf{y} e^{-\beta H_{SB}(\mathbf{x},\mathbf{y})} \rho_B(\mathbf{y}, \beta)\:,
\label{eH*cl}
\end{equation}   
where $\mathbf{x}$ and $\mathbf{y}$ denote points in the system and the environmental phase-space, respectively, $d \mathbf{y}$ the dimensionless phase-space volume element, and
\begin{equation}
\rho_B(\mathbf{y},\beta) = Z^{-1}_B(\beta) e^{-\beta H_B(\mathbf{y})}
\label{rB}
\end{equation} 
the canonical phase-space pdf of the bare environment, i.e., in the absence of the system. Hence, the classical Hamiltonian of mean force can be expressed as
\begin{equation}
H^*(\mathbf{x},\beta,\lambda) = H_S(\lambda) - \beta^{-1} \ln \langle e^{-\beta H_{SB}(\mathbf{x},\mathbf{y})} \rangle_B\:,
\label{H*cl}
\end{equation}
with $\langle \bullet \rangle_B = \int d\mathbf{y} \bullet \rho_B(\mathbf{y},\beta)$ standing for the average over the canonical state of the bare environment. Note that in the classical case the renormalization is determined by the second term on the right-hand side of Eq. (\ref{H*cl}), which is independent of the system parameters $\lambda$.
For the typical case of a system that solely couples the positions of the system $\mathbf{Q}$ and the environment $\mathbf{q}$   
via the potential $V_{\text{SB}}(Q,\mathbf{q})$
only the scalar potential of the system is renormalized. The resulting total potential $V^*(\mathbf{Q})$ is then known as the potential of mean force~\cite{Kirkwood35JCP,Hanggi90RMP,Onsager33}. It is given by
\begin{equation}  
V^*(\mathbf{Q}) = V(\mathbf{Q}) - \beta^{-1} \ln \langle e^{-\beta         
 V_{\text{SB}}(\mathbf{Q},\mathbf{q}) } \rangle_B\:,
\label{V*}
\end{equation}
where $V(\mathbf{Q})$ denotes the potential of the bare system.

For later use we note the  identity 
\begin{equation}
\frac{\partial}{\partial \beta} \left [ \beta H^*(\mathbf{x},\beta,\lambda) \right ] = \langle H_{\text{tot}}|\mathbf{x} \rangle - \langle H_B \rangle_B
\label{H*Htot}
\end{equation}
relating the Hamiltonian of mean force and its inverse temperature derivative to the deviation of the average of the total Hamiltonian conditioned on the state of the system ${\bf x}$ from the average of the bare environmental Hamiltonian~\cite{Talkner16bPRE}.
The average $\langle \bullet | \mathbf{x} \rangle = \int d \mathbf{y} \bullet w(\mathbf{y}|\mathbf{x}) $ is performed with respect to 
the conditional probability density function (pdf) $w(\mathbf{y}|\mathbf{x})$ of finding the environment at the phase-space point $\mathbf{y}$ once the system is at $\mathbf{x}$. As such it is given by 
\begin{equation}
\begin{split}
w(\mathbf{y}|\mathbf{x})& =\frac{\rho_{\text{tot}}(\mathbf{x},\mathbf{y},\beta,\lambda)}{\rho_S(\mathbf{x},\beta,\lambda)}\\
&=Z_B^{-1} e^{-\beta \bb {H_{\text{tot}}(\mathbf{x},\mathbf{y},\lambda) -H^*(\mathbf{x},\beta,\lambda)}} \:.
\end{split}
\label{wyx}
\end{equation} 
This conditional pdf characterizes the equilibrium preparation class of open classical systems~\cite{Grabert77ZPB}. It plays a key role in the projection operator formulation of the open system's dynamics~\cite{Zwanzig61PR,Grabert80JSP, Grabert82Book}.

In the case of weak coupling~\cite{vanHove57Phys,Davies76Book} the equilibration of the considered system with its environment is achieved  by a  vanishingly small interaction strength $\kappa \to 0$ at correspondingly late times $t \to \infty$ with $\kappa^2 t$ finite. Accordingly, this weak interaction does not cause a renormalization of the system Hamiltonian and hence the Hamiltonian of mean force coincides with the Hamiltonian of the bare system.  For a further discussion of the weak coupling limit see  Sec.~\ref{weakcoupling}. 
\subsection{Thermodynamics}\label{Thermo}
The thermodynamics of an open system being part of a canonical total system follows  from its partition function [Eq.(\ref{ZSZtotZB})] in the standard way starting with the Helmholtz free energy given by
\begin{equation}
F_S(\beta,\lambda) = - \beta^{-1} \ln Z_S(\beta,\lambda)\:.
\label{FS}
\end{equation}
Likewise, one obtains the internal energy and the entropy of the open system by means of the thermodynamic relations (\ref{U}) and (\ref{S}). As a consequence of the open system partition function being the ratio of the total system's and the bare bath' functions according to Eq. (\ref{ZSZtotZB}), all thermodynamic potentials and also all response functions result as differences of the respective quantities of the total system and the bare bath. Therefore, they are of the form
\begin{equation}
\Xi_S =\Xi_{\text{tot}} -\Xi_B\:,
\label{XiXi}
\end{equation}
where $\Xi$ stands for any of the thermodynamic potentials. Moreover, $\Xi$ may stand for the specific heat  $C = \partial U/\partial T|_\lambda$, and
the response to variations of local or global parameters such as the isothermal magnetization $M_T = \partial F/ \partial \sf{h}|_T$ given by the derivative of the free energy with respect to an external magnetic field $\sf{h}$, isobaric thermal expansion coefficients, as well as for all susceptibilities specified by the second derivatives of the thermodynamic potentials with respect to the relevant, local or global parameters of the total system, such as the isothermal magnetic susceptibility $\chi = \partial^2 F/ \partial \sf{h}^2|_T$. 
Here, the indices $S$, $\text{tot}$ and $B$ stand for open system, total system, and bare bath, respectively.  
This particular form also ensures the validity of the third law of thermodynamics provided that the individual entropies of the total system and the environment vanish when the temperature approaches the absolute zero point~\cite{Hanggi06APPB}. As a difference of two quantities, at finite coupling strength, the entropy as well as the specific heat and, at finite coupling, the susceptibilities need {\it not} comply with their standard positivity properties: The entropy and the specific heat may become negative in certain parameter regions and the susceptibility matrix may also violate positivity.  Examples are given later.

Finally, we note that the construction of the thermodynamic potentials of an open system is not restricted to canonical states of the total system but rather can be extended to pressure  or grand canonical ensembles with fluctuating volume or particle number, respectively. For example, with the replacement of the Boltzmann factor $e^{-\beta H_{\text{tot}}(V)}$ of the total system at fixed volume $V$ with $\int dV e^{-\beta \bb {H_{\text{tot}}(V) + p V  }}$,  allowing for volume fluctuations controlled by the external pressure $p$ one obtains in an analogous way  for the Hamiltonian of mean force  
\begin{equation}
H^*(\beta,p) =-\beta^{-1} \ln \frac{ \int dV \Tr_B e^{-\beta \bb {H_{\text{tot}}(V) + p V } }}{  \int dV \Tr_B e^{-\beta \bb { H_B(V) + p V }  }}
\label{H*p}
\end{equation}
 and accordingly for the pressure-dependent partition function of the open system one obtains $Z_S(\beta,p) = Z_{\text{tot}}(\beta,p)/Z_B(\beta,p)$, where $Z_X(\beta,p) = \int dV \Tr_X e^{-\beta \bb {H_X(V) +p V }}$ for $X= \text{tot},\:B$. A possible dependence on further parameters $\lambda$ has been suppressed so as not to overburden the notation. All thermodynamic potentials, such as the enthalpy and entropy as well as their derivatives, 
are obtained from the system Gibbs free energy given by $G_S(\beta,p) = -\beta^{-1} \ln Z_S(\beta,p)$. Consequently, they are 
again  determined as the differences of the respective functions of the total  system and the bare bath, guaranteeing thermodynamic consistency.~\footnote{In the so-called bare representation, suggested by~\citet{Jarzynski17PRX}, the entropy is given by the  Shannon entropy of the reduced open system phase-space pdf, and hence the bare representation violates thermodynamic consistency in general beyond the weak coupling regime.}         
\subsection{Statistical mechanical expressions of the thermodynamic potentials}\label{statmechpot}
The statistical mechanical expression for the free energy of an open system with strong coupling is of the same formal structure as for weak coupling with the  Hamiltonian of mean force replacing the bare Hamiltonian. For the free energy $F_S(\beta,\lambda)$ hence one  obtains from eqs. (\ref{ZSZtotZB}) and (\ref{FS}) that
\begin{equation}
F_S(\beta,\lambda) = - \beta^{-1} \ln \Tr e^{-\beta H^*(\beta,\lambda)}\:.   
\label{FSH*}
\end{equation}
When going from the free energy to the internal energy by means of Eqs.~(\ref{U}, \ref{FS}, \ref{FSH*}) one finds
\begin{equation}
U_S(\beta, \lambda) = \langle H^*(\beta,\lambda) \rangle_S + \beta \Big \langle \frac{\partial H^*(\beta,\lambda)}{\partial \beta} \Big \rangle_S\:,     
\label{USH*}
\end{equation}
where $\langle \bullet \rangle_S = \Tr_S \bullet \rho_S(\beta,\lambda)$ denotes the average with respect to the equilibrium system density matrix $\rho_S(\beta,\lambda) = Z_S^{-1} e^{-\beta H^*(\beta,\lambda)}$ in the quantum case. In the classical case, the trace must be replaced by the respective phase-space integral and the density matrix by the respective phase-space pdf.  The first term on the right hand side corresponds to the standard expression of the internal energy in terms of the equilibrium average of the system Hamiltonian. The second term, which is specific for open systems interacting with their environments at a nonvanishing strength, is a direct consequence of the temperature dependence of the Hamiltonian of mean force. Expressing the internal system energy with Eqs. (\ref{Htot} and \ref{XiXi}) as the difference of the internal energies of the total system and the bare bath, one obtains an alternative expression of the form~\cite{Hanggi08NJP,Hanggi06APPB}
\begin{equation}
U_S(\beta,\lambda) = \langle H_S(\lambda) \rangle_S + \langle H_{SB} \rangle_{\text{tot}} + \langle H_B \rangle_{\text{tot}} - \langle H_B \rangle_B\:.
\label{USHS}
\end{equation}  
Here $\langle \bullet \rangle_{\text{tot}} = \Tr_{SB} \bullet \rho_{\text{tot}}(\beta,\lambda)$ with the trace over the product Hilbert space of the system and the environment $\Tr_{SB}=\Tr_S \Tr_B$ indicates an average over the canonical state of the total system, and, as previously defined, $\langle \bullet \rangle_B = \Tr_B \bullet \rho_B(\beta)$ is the average over the canonical state of the bare environment $\rho_B(\beta) = Z^{-1}_B(\beta) e^{-\beta H_B}$. 
The last three terms on the right-hand side of Eq.~(\ref{USHS}) describe the deviation of the internal energy from the average Hamiltonian of the bare system due to the interaction induced system-environment correlations. They become negligible in the aforementioned weak coupling limit and also in exceptional cases such as for classical open systems with a single degree of freedom in contact with a Caldeira-Leggett-type heat bath~\cite{Ford65JMP,Zwanzig73JSP,Caldeira83AP,Hanggi90RMP,Magalinskii59JETP,Bogolyubov45PAS}. Note that in spite of the $\lambda$ independence of the Hamiltonians $H_{SB}$ and $H_B$ their averages with respect to the total equilibrium density matrix $\rho_{\text{tot}}$ do, in general, depend on $\lambda$. Neither these terms nor the internal energy of the bare bath  must  be neglected, or are only partly taken into account, as one may find in the literature~\cite{Sekimoto98PTP,Seifert12RPP,Strasberg17PRX,Hsiang18PRE,Dou18PRB,Jarzynski17PRX}.~\footnote{The argument stated by \citet{Seifert19ARCMP} and \citet{Strasberg20arX}, that only differences of the same thermodynamic quantities at the beginning and the end of a thermodynamic process are of relevance, restricts the applicability of the corresponding approach to isothermal processes at constant external parameters. See also the discussion by \citet{Talkner20arX}.}

For the entropy of an open system one finds from Eq. (\ref{S}) in combination with (\ref{ZSZtotZB}) and (\ref{FS})  the expression
\begin{equation}
\begin{split}
S_S(\beta,\lambda) & = k_B \frac{\partial T \ln Z_S(\beta,\lambda)}{\partial T}\\
&= k_B \ln Z_S - k_B \beta \frac{\partial}{\partial \beta} \ln Z_S\\
&= \mathfrak{S}\bb {\rho_S(\beta,\lambda) } + k_B \beta^2 \Big \langle \frac{\partial H^*(\beta,\lambda)}{\partial \beta} \Big \rangle_S\:.
\end{split}
\label{SSH*}
\end{equation}
Here $\mathfrak{S}(\rho)$ denotes the von Neumann entropy of the density matrix $\rho$, $\mathfrak{S}(\rho) = - k_B \Tr \rho \ln \rho$ for quantum systems~\cite{vonNeumann55Book}  and for classical systems the respective continuous Shannon entropy 
$\mathfrak{S}(\rho) = - k_B \int d \mathbf{x} \rho(\mathbf{x}) \ln \rho(\mathbf{x})$ of the phase space pdf with respect to a properly defined dimensionless infinitesimal phase space volume $d \mathbf{x}$ ~\cite{Wehrl78RMP}.~\footnote{While the von Neumann entropy remains unchanged under unitary transformations, and likewise the  Shannon entropy  is invariant under canonical transformations of  phase-space variables, the latter is not invariant under general transformations ~\cite{Marsh13}. This is so because $\rho(\mathbf{x})$ transforms under any transformation $\mathbf{x} \to \mathbf{y} = \mathbf{f}(\mathbf{x})$ as a density, i.e. $\bar{\rho}(\mathbf{y}) = \rho(\mathbf{f}^{-1}(\mathbf{y}))/| d \mathbf{y}/d\mathbf{x}|$ and, due to the presence of the Jacobian $|d \mathbf{y}/d\mathbf{x}|$, not as a scalar function. This renders the Shannon entropy of a marginal pdf, as for the configuration space or a part of it, an ill defined expression.}  
 As for the internal energy of open systems outside the weak coupling regime,  an additional contribution to the standard von Neumann-Shannon form of  entropy exists, in general, due to the temperature dependence of the Hamiltonian of mean force. This guaranties thermodynamic consistency, which means that the free and internal energy and the entropy are related to each other in the standard way expressed in Eq.~(\ref{FUS}). The presence of the extra term though leads to the fact that the thermodynamic entropy of an open system deviates from the information entropy according to Shannon.   
Because the Hamiltonian of mean force $H^*(\beta,\lambda)$ is not uniquely determined by the reduced density matrix $\rho_S(\beta,\lambda)$, the thermodynamic entropy cannot be expressed as a functional of the reduced state of the open system alone. 

The entropy of an open system can likewise be expressed in terms of the Kullback Leibler divergence also known as the relative entropy,~\cite{Wehrl78RMP} $S \bb {\rho_{\text{tot}}(\beta,\lambda)||\rho_S(\beta,\lambda) \otimes \rho^B(\beta,\lambda)}$ between the total state $\rho_{\text{tot}}(\beta,\lambda)$ and the product state $\rho_S(\beta,\lambda) \otimes \rho^B(\beta,\lambda)$ and von Neumann entropies of the system and the environment, given by
\begin{equation}
\begin{split}
S_S(\beta,\lambda) &= -S \bb {\rho_{\text{tot}}(\beta,\lambda)||\rho_S(\beta,\lambda) \!\otimes\! \rho^B(\beta,\lambda) }\\
&\quad +\mathfrak{S}\bb {\rho_S(\beta,\lambda) }\\
& \quad +
\mathfrak{S}\bb {\rho^B(\beta,\lambda) } - \mathfrak{S}\bb {\rho_B(\beta) }\:.
\end{split}
\label{KLS}
\end{equation}
Here the relative entropy of the density matrices (phase-space pdf's) $\rho$ and $\tau$ is defined as $S(\rho||\tau) =\Tr \rho ( \ln \rho - \ln \tau)$, and the reduced environmental state $\rho^B(\beta,\lambda)$ is given by
\begin{equation}
\rho^B(\beta,\lambda) = \Tr_S \rho_{\text{tot}}(\beta,\lambda)\:.
\label{rBrtot}
\end{equation} 
Hence, the contribution to the system entropy 0n the last line of Eq.~(\ref{KLS}) comes from the difference of the von Neumann entropies of the reduced and the bare environmental state, $\rho^B(\beta,\lambda)$ and $\rho_B(\beta)$, respectively; this difference vanishes in the weak coupling limit. In addition, in the latter limit the relative entropy vanishes and, as expected, the entropy of the open system agrees with the von Neumann entropy. 

The open system entropy can also be expressed in terms of the conditional entropy of the {\it system}, given the state of the environment, $S(S|B) = \mathfrak{S}\bb {\rho_{\text{tot}}(\beta,\lambda) } -\mathfrak{S}\bb {\rho^B(\beta,\lambda) } $ to read
\begin{equation}
S_S(\beta,\lambda) = S(S|B) + \mathfrak{S}\bb {\rho^B(\beta,\lambda) } - \mathfrak{S}\big (\rho_B(\beta) \big )\:.
\label{SSSB}
\end{equation}
Similarly one may also express the open system entropy $S_S(\beta,\lambda)$ in terms of the Bayesian sibling $S(B|S) = \mathfrak{S}(\rho_{\text{tot}}(\beta,\lambda) - \mathfrak{S}(\rho_S(\beta,\lambda) $ as
\begin{equation}
S_S(\beta,\lambda) = S(B|S) + \mathfrak{S}\bb {\rho_S(\beta,\lambda) } - \mathfrak{S}\bb {\rho_B(\beta) } \:.
\label{SSBS}
\end{equation}
In passing, we note that for classical systems the conditional entropy of the {\it bath} can be written in terms of the conditional pdf $w(\mathbf{y}|\mathbf{x})$ characterizing the environmental phase-space distribution once the system state $\mathbf{x}$ of the open system is specified.  
\begin{equation}
S(B|S) = - k_B \int d \mathbf{x} d \mathbf{y} \rho_{\text{tot}}(\mathbf{x},\mathbf{y},\beta,\lambda) \ln w(\mathbf{y}|\mathbf{x})\:,
\label{SBS}
\end{equation}
where the conditional pdf $w(\mathbf{y}|\mathbf{x})$ is given by Eq.~(\ref{wyx}).

While the von Neumann entropy is always positive, none of the equations (\ref{XiXi}), (\ref{SSH*}), or (\ref{SSSB}) guarantee that the entropy of an open system may not take negative values.

Similarly, in view of Eq. (\ref{XiXi}), the specific heat $C_S$ and the susceptibilities $\chi_S$ may assume negative values without indicating any instability of the considered open system~\cite{Ingold09PRE,Campisi09JPA,Campisi10CP}. 
The specific heat can be equally expressed in terms of the partition function as
\begin{equation}
C_S = k_B \beta^2 \frac{\partial^2 \ln Z_S}{\partial \beta^2}\:.
\label{CSZS}
\end{equation}    
Consequently, in view of the possibility of a negative specific heat, the internal system energy $U_S = \partial \ln Z_S /\partial \beta$ may  decrease with increasing temperature. 

\subsection{Examples}\label{Exa}
\begin{figure*}[t]
\includegraphics[width=0.6\columnwidth]{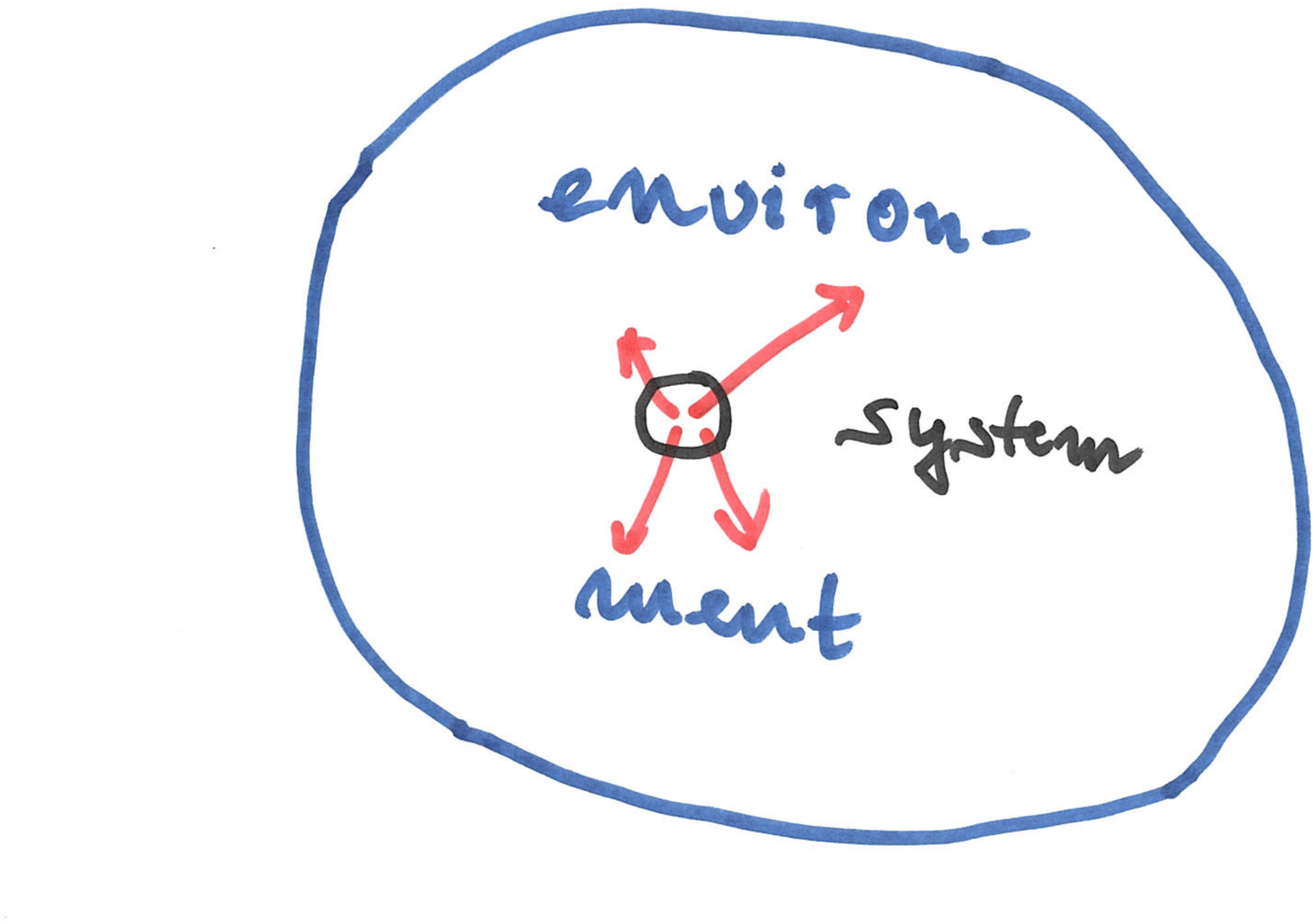}
\includegraphics[width=0.6\columnwidth]{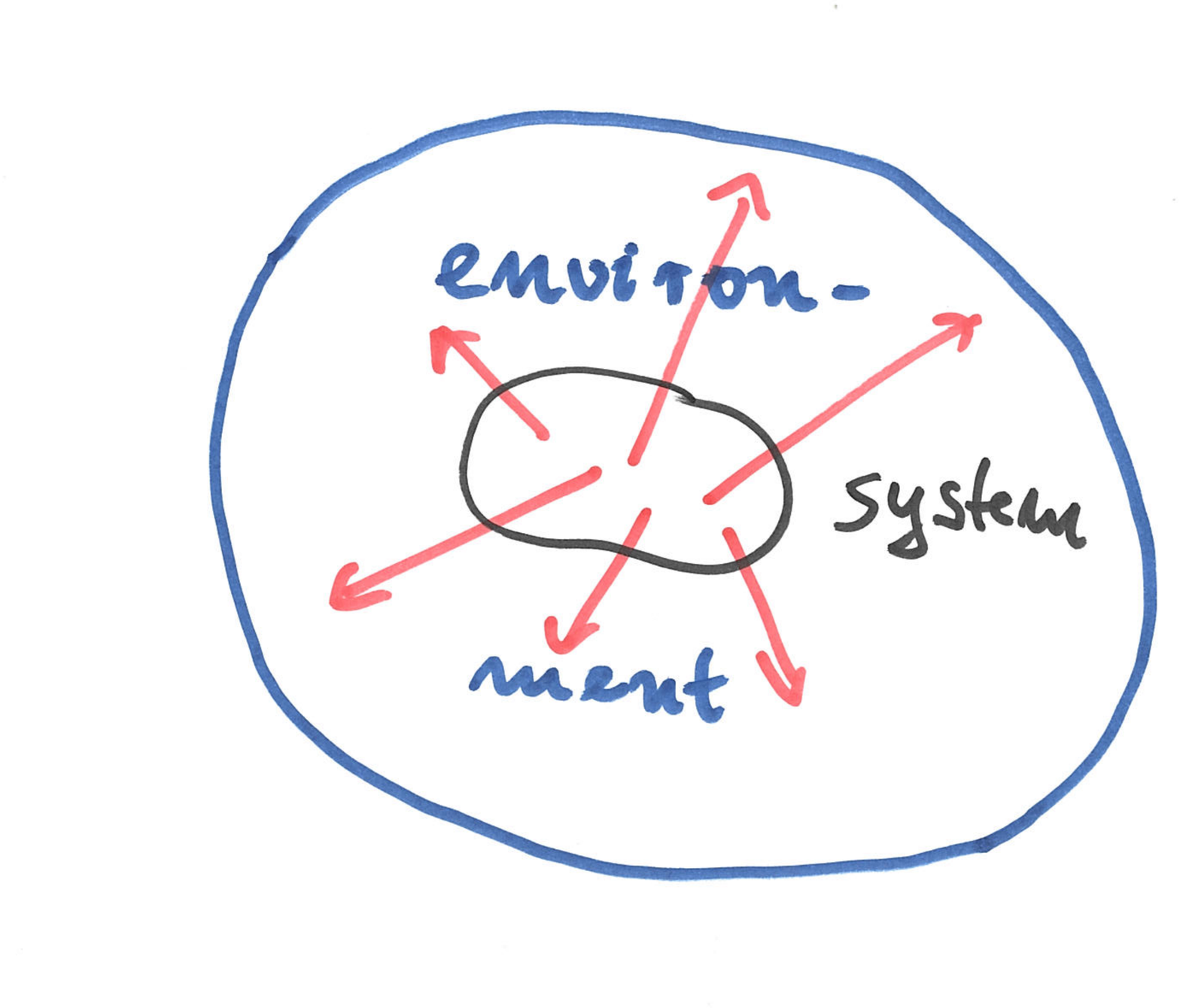}
\includegraphics[width=0.6\columnwidth]{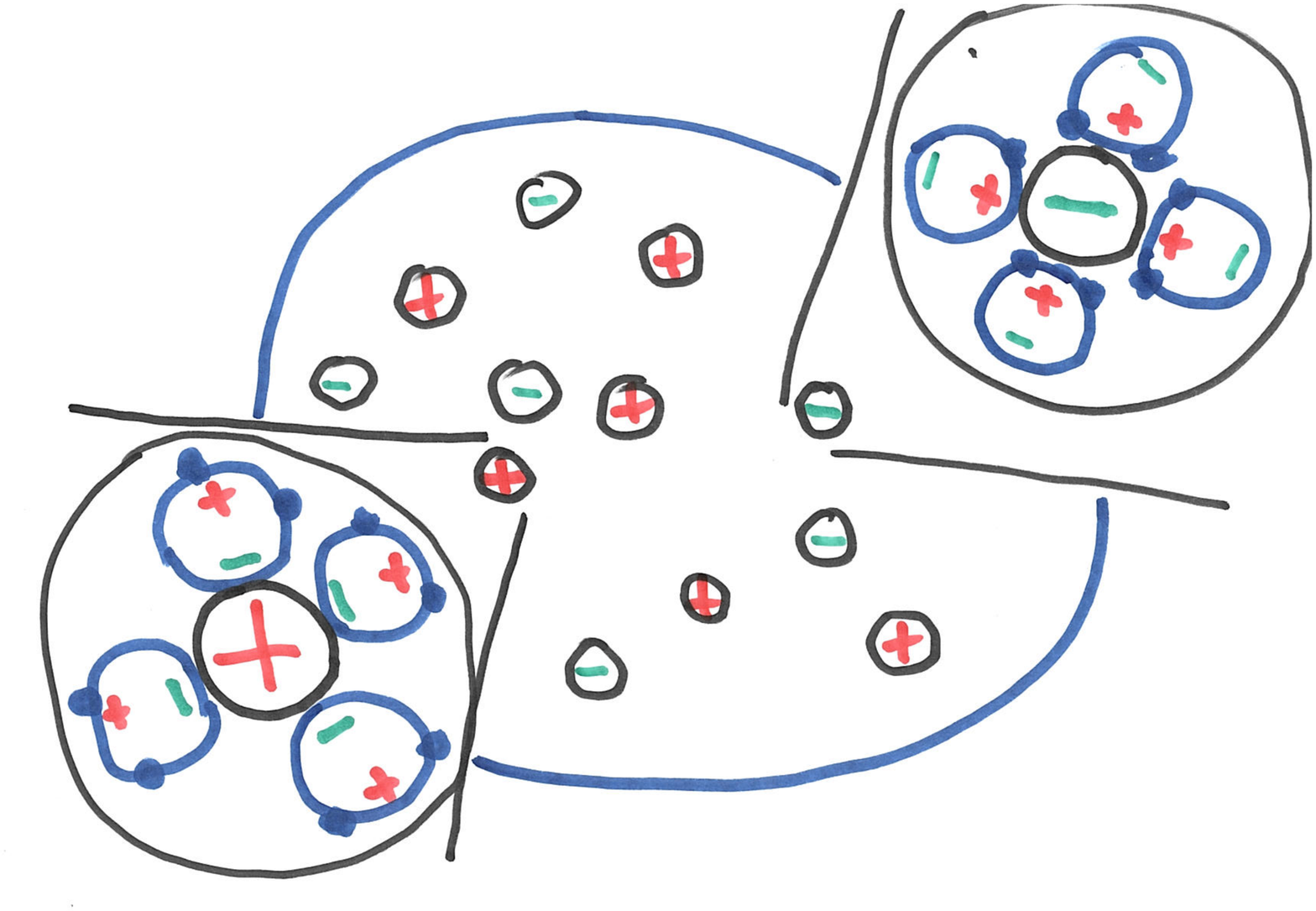}
\caption
{Different realizations of system-environment coupling.  Left panel:  a single microscopic entity interacting with the neighborhood of its environment. Middle panel: all parts of a mesoscopic system interacting with at least a part of the environment. This strong coupling situation persists for a macroscopic system if the interaction forces are long-range. Right panel:   solute in a solvent such as salt in water. Typically, the dipolar water molecules orient themselves in the vicinity of an ion as a way to shield its charge, as sketched in the magnification. 
}
\label{F1}
\end{figure*} 

Before we consider specific examples, we  classify those  situations in which an open system couples to its environment with a strength beyond the weak coupling limit. As indicated in  Fig.~\ref{F1}, we may distinguish roughly three scenarios. In the first one, sketched  in the  left panel,  a single microscopic object with a few degrees of freedom interacts with its environment; the immediate interaction may be restricted to the close vicinity of this object but may also extend out into more distant parts of the environment. In any case, the weak coupling limit is reached only if, in thermal equilibrium, the interaction energy is negligible compared to the energy of the system degrees of freedom. The second class, sketched in the middle  panel of Fig.~\ref{F1}, comprises open systems of mesoscopic, or even macroscopic, size that can be distinguished by their shape and physical properties from the surrounding environment. Here the weak coupling regime sets in when the interaction potential is short range with a characteristic length that is much shorter than the typical linear dimension of the system. Deviations from weak coupling appear in cases of strong short-range interactions and also in relatively weak long-range interactions. The third scenario, pictured in the right panel, is essential for the theory of solutions~\cite{Roux99BCP} and can be exemplified by a grain of salt dissolved in a glass of water. As in the second scenario the amount of substance constituting the open system may vary from mesoscopic, say, the size of a cluster of a few hundred atoms, to macroscopic. However, in contrast to the previous case, many of the physical properties including the spatial extension may change fundamentally. In this scenario the weak coupling limit is hardly  ever reached. The solution scenario evidently illustrates the influence of the specific properties of the environment on those of the open system: The same solute may behave quite differently in different solvents.            

\subsubsection{Damped harmonic oscillator}\label{Dho}
A damped harmonic oscillator in contact with a heat bath at the inverse temperature $\beta$ can be described in terms of the Zwanzig-Caldeira-Leggett model~\cite{Zwanzig73JSP,Caldeira83AP}, where an oscillator of mass $M$, position $Q$ and momentum $P$ couples to a bath of many other harmonic oscillators mimicking the environment.   The Hamiltonian of  the total system is given by
\begin{equation}
\begin{split}
H_{\text{tot}} &= \frac{P^2}{2M} +\frac{1}{2} M \Omega^2 Q^2\\
&\quad + \sum_n \left \{ \frac{p^2_n}{2 m_n} + \frac{1}{2}m_n \omega^2_n \left (q_n - \frac{C_n}{m_n \omega^2_n} Q \right )^2 \right \}\:.
\end{split}
\label{ZCL}
\end{equation}
where $\Omega$ is the frequency of the uncoupled oscillator, while $m_n$, $\omega_n$, $q_n$ and $p_n$ are the mass, frequency, position, and momentum of the $n$th bath oscillator. The parameters $C_n$ determine the coupling strength between the system oscillator and the $n$th bath oscillator.  
For the behavior of the central oscillator,  with respect to both its dynamical and equilibrium properties, it is sufficient to specify the so-called memory kernel~\cite{Grabert88PR,Weiss08Book}
\begin{equation}
\gamma(t)= \frac{1}{M}\sum_n \frac{C^2_n}{m_n \omega^2_n} \cos \omega_n t\:.
\label{gt}
\end{equation}
The partition function $Z_S(\beta,\Omega)$ can be expressed in terms of the Laplace transform of the memory kernel, $\hat{\gamma}(s) = \int_0^\infty e^{-s t} \gamma(t)$, as an infinite product of the form~\cite{Weiss08Book}
\begin{equation}
Z_S(\beta, \Omega) = \frac{1}{\beta \hbar \Omega} \prod_{n=1}^\infty \frac{\nu^2_n}{\Omega^2 + \nu^2_n +\nu_n \hat{\gamma}(\nu_n)}\:,
\label{ZSosc}
\end{equation}
where $\nu_n =2 \pi n/(\hbar \beta)$ and $n=1,2,\ldots$ denote the Matsubara frequencies.
The free energy resulting from Eqs.  (\ref{FS}) and (\ref{ZSosc}) agrees with Ford, Lewis, and O'Connell's ``remarkable formula''~\cite{Ford85PRL} expressing $F_S(\beta,\Omega)$ as
\begin{equation}
F_S(\beta,\Omega) = \frac{1}{\pi} \int_0^\infty d\omega f(\beta,\omega) \text{Im}\left ( \frac{d \ln \chi(\omega)}{d \omega} \right )\:.
\label{FSosc}  
\end{equation}
Here, $f(\beta,\omega)= \beta^{-1} \ln (2 \sinh \beta \hbar \omega/2) $ is the free energy of  an isolated  harmonic oscillator with frequency $\omega$ and $\chi(\omega) = 1/[M(\Omega^2 -\omega^2) - i \omega \hat{\gamma} (i \omega)]$ denotes the susceptibility of the damped oscillator.
 
For the Drude model, which is specified by $\gamma(t) = \gamma \omega_D e^{-\omega_D t}$ with the Drude-frequency $\omega_D$ and the static damping constant $\gamma$, the partition function of the oscillator can be expressed in closed form \cite{Grabert84ZPB} as
\begin{equation}
Z_S(\beta,\Omega) = \frac{\beta \hbar \Omega}{4 \pi^2 } \frac{\Gamma(\lambda_1/\nu)\Gamma(\lambda_2/\nu)\Gamma(\lambda_3/\nu)}{\Gamma(\omega_D/\nu)}\:,
\label{ZSoscD}
\end{equation}  
where $\Gamma(z)$ denotes the gamma function~\cite{Abramowitz64Book}, $\lambda_i$, $i=1,2,3$ denotes the solutions of the cubic equation $\lambda^3 - \omega_D \lambda^2 +(\Omega^2+ \gamma \omega_D) \lambda - \omega_D \Omega^2 =0$, and $\nu= \nu_1$ denotes the fundamental Matsubara frequency.
The first moments $\langle Q \rangle$ and $\langle P \rangle$ of position and momentum, respectively, vanish; the second moments can be expressed in terms of logarithmic derivatives of the partition function yielding
\begin{align}
\langle Q^2 \rangle &= -\frac{1}{M} \beta \Omega \frac{\partial \ln Z_S}{\partial \Omega}\:,
\label{Q2}\\
\langle P Q\rangle&= \frac{\hbar}{2i}\:,
\label{PQ}\\
\langle P^2 \rangle &= M^2 \Omega^2 \langle Q^2 \rangle - \frac{2 M \gamma}{\beta} \frac{\partial \ln Z_S}{\partial \gamma}\:. 
\label{P2}
\end{align}
Note that the symmetrized position-momentum correlation function $\langle P Q + Q P \rangle$ vanishes because of the time-reversal invariance of the thermal equilibrium state. 
Because the state of the total system is Gaussian, the reduced density matrix is of Gaussian form and hence is completely determined by its first two moments \cite{Talkner81ZPB}.
The reduced density matrix of the damped oscillator then becomes
\begin{equation}
\rho_S = Z^{-1}_{\text{eff}} e^{-\beta H_{\text{eff}}}\:,
\label{rhoSosc}
\end{equation}
where  $Z_{\text{eff}} =2 \sinh (\beta \hbar \Omega_{\text{eff}}/2)$ and the effective Hamiltonian is quadratic in position and momentum, given by
\begin{equation}
H_{\text{eff}} = \frac{1}{2 M_{\text{eff}}} P^2 +\frac{1}{2} M_{\text{eff}} \Omega^2_{\text{eff}} Q^2\:.
\label{Heff}
\end{equation} 
The renormalized frequency and mass can be expressed as~\cite{Grabert84ZPB}
\begin{align} 
 \Omega_{\text{eff}}&= \frac{1}{\hbar \beta} \ln \frac{\sqrt{\langle P^2\rangle \langle Q^2 \rangle}+\hbar/2}{\sqrt{\langle P^2\rangle \langle Q^2 \rangle}-\hbar/2} \:,
\label{Oeff}\\
M_{\text{eff}} &=\frac{1}{\Omega_{\text{eff}}}\sqrt{\frac{\langle P^2 \rangle}{\langle Q^2 \rangle}}\:.
\label{Meff} 
\end{align}
Note that the Hamiltonian of mean force does not coincide with  $H_{\text{eff}}$, despite other claims~\cite{Philbin16JPA,Miller18}, because the resulting normalizing effective partition function $Z_{\text{eff}}=\Tr_S e^{-\beta H_{\text{eff}}} = [2\sinh(\beta \hbar \Omega_{\text{eff}}/2)]^{-1}$ does not agree with the open system partition function $Z_S$ that is given by  Eq.~(\ref{ZSoscD}). The Hamiltonian of mean force is instead given by 
\begin{equation}
H^* = H_{\text{eff}} + \beta^{-1} \ln (Z_{\text{eff}}/Z_S)\:.
\label{H*osc}
\end{equation}
For small temperatures, $\beta \hbar \Omega_{\text{eff}}$ approaches a finite value, depending on $\gamma/\Omega$ and $\omega_D/\Omega$, with the consequence that the von Neumann entropy $\mathfrak{S}(\rho_S)/k_B= \beta \hbar \Omega_{\text{eff}}/2 \coth \beta \hbar \Omega_{\text{eff}}/2 - 2 \ln [2 \sinh (\beta \hbar \Omega_{\text{eff}}/2) ]$ of the oscillator converges in this limit to a value other than zero, indicating an entanglement between the oscillator and its environment~\cite{Horhammer08JSP} in the ground-state wave function of the total system.  In contrast to the von Neumann entropy $\mathfrak{S}(\rho_S)$ the thermodynamic entropy vanishes at low temperatures in agreement with the third law of thermodynamics~\cite{Hanggi06APPB,Hanggi08NJP,Ingold09PRE}.

The classical limit for the damped harmonic oscillator can be performed by letting the dimensionless parameter $\beta \hbar \Omega$ approach zero, yielding for the partition function $Z_S$ the classical value $Z^{\text{cl}}_S = 1/(\beta \hbar \Omega)$ of a bare harmonic oscillator in a canonical state at the inverse temperature $\beta$. 
Likewise, the Hamiltonian of the mean force approaches in the classical limit the Hamiltonian of the bare oscillator independent of the interaction strength. This must not be seen as specifically classical behavior but instead as a peculiarity  of the Zwanzig-Caldeira-Leggett model and related models for systems with a single degree of freedom.\footnote{\label{fn_canonical}For any total Hamiltonian with a potential function $ U(\{q \},Q) = V(Q) + F(\{ q_n - g_n Q \}) $ jointly describing the 
potential energy of the system and of the environmental degrees of freedom as well as the interaction between system and environment, the integral in the Eq. (\ref{eH*cl}) representing the renormalized Boltzmann factor can be performed with the help of the  coordinate transformation $q_n - g_n Q \to \bar{q}_n$ for all $n$, yielding unity such that Eq.~(\ref{V*}) results in $V^*(Q)=V(Q)$. }  The modeling of the environment according to~\citet{Ullersma66Physica} differs from the Zwanzig-Caldeira-Leggett model by the absence of the ``counter term'' $\sum_k C^2_k /(2 m_k \omega^2_k) Q^2$ in the total Hamiltonian. This term  then appears with the opposite sign in the classical Hamiltonian of mean force.~\footnote{\label{fn_stability}For a harmonic oscillator this frequency renormalization may render the system unstable, restricting the choice of the possible environmental parameters. No restrictions of this kind exist for systems with potentials that are more strongly repulsive at infinity than harmonic ones.} 

Another example displaying a nontrivial Hamiltonian of mean force is given by two particles coupling to the same Zwanzig-Caldeira-Leggett-type environment~\cite{Duarte06PRL,Valente10PRA}. To be specific, we consider a classical system described by the total Hamiltonian
\begin{equation}
\begin{split}
H_{\text{tot}}&=H_S+\sum_{n=1}^N \Bigg [\frac{p^2_n}{2 m_n} \Big .\\
&\quad  \Big . + \frac{1}{4}  m_n \omega^2_n \sum_{k=1,2} \left ( q_n - \frac{2 C_{n,k}}{m_n \omega^2_n}   Q_k \right )^2 \Bigg ]\:,
\end{split}
\label{H2tot}
\end{equation} 
where $H_S$ is the Hamiltonian of the isolated system with 2 degrees of freedom specified by the coordinates $Q_1$ and $Q_2$ which both couple to the coordinates $q_n$ of $N$ harmonic oscillators with coupling strength $C_{n,k}$, with $k=1,2$ labeling the system's degrees of freedom. The Hamiltonian of mean force can be calculated by performing Gaussian integrals over the environment degrees of freedom yielding
\begin{equation}
H^*= H_S + \delta V^*\:,
\label{H2*}
\end{equation} 
where the potential renormalization is given by
\begin{equation}
\delta V^* =\sum_n \frac{(C_{n,1} Q_1 + C_{n,2} Q_2)^2}{2 m_n \omega^2_n}\:.
\label{V2*}
\end{equation}
This potential causes in general environment induced forces, both on the center of mass $Q=(M_1 Q_1 +M_2 Q_2)/(M_1+M_2)$ and on the relative coordinate $x =Q_1 -Q_2$. The force on the center of mass vanishes if the total interaction constants are equal, i.e. for $\sum_n C_{n,1}/(2m_n \omega^2_n) =\sum_n C_{n,2}/(2m_n \omega^2_n)$. The force acting on the relative coordinate can be either attractive or repulsive, depending on the parameter values.
Because of the temperature {\it independence} of the potential of mean force the thermodynamic entropy of the open system $S_S$, as specified in Eq. (\ref{SSH*}), coincides with the Gibbs-Shannon entropy $\mathfrak{S}(\rho_S)$ of the reduced system pdf $\rho_S = Z^{-1}_S e^{-\beta H^*}$. The partition function $Z_S$ of the open system coincides with the ratio of the total system and the bare environment in accordance with Eq.~(\ref{ZSZtotZB}).   
\subsubsection{Damped free particle}\label{Dfp}
In a manner similar  to the case of a damped harmonic oscillator, a damped free particle of mass $M$ can be modeled by means of the Zwanzig-Caldeira-Leggett Hamiltonian (\ref{ZCL}) by disregarding the parabolic system potential, that is, by setting $\Omega =0$~\cite{Grabert88PR}. To prevent the particle from escaping  to infinity and  guarantee the existence of a  normalizable thermodynamic equilibrium state, a confining box of large length $L$ is introduced.  
In spite of its seeming simplicity, an exact expression for the partition function of the total system and consequently also for the free damped free particle is not known. Only for sufficiently high temperatures, for which the Gauss sum representing the partition function $Z^0_S= \sum_n e^{-\beta E_g n^2}$ of the bare system  converges to the respective Gaussian integral,\footnote{\label{fn_convergence}A convergence of the Gauss sum to the classical free particle partition function better than 1\% is achieved for $\beta E_g=\beta \pi^2 \hbar^2/(2 M L^2) \lessapprox 10^{-4}$.} can the partition function $Z_S$ of the damped particle be approximated by the following expression~\cite{Hanggi08NJP}: 
\begin{equation}
Z_S= \frac{L}{\hbar} \left (\frac{2 \pi M}{\beta} \right )^{1/2} \frac{\Gamma(1+x_1) \Gamma(1+x_2)}{\Gamma(1+\hbar \beta \omega_D/(2 \pi))}\:,
\label{ZSfp}
\end{equation}
where 
\begin{equation}
x_{1,2} = \frac{\hbar \beta \omega_D}{4 \pi} \left (1 \pm \sqrt{1 -4 \gamma/\omega_D} \right )\:.
\label{x12}
\end{equation}       
Here $E_g = \pi^2 \hbar^2/(2M L^2)$ denotes the ground-state energy of the  free particle in a box of length $L$.   
Based on the approximate expression (\ref{ZSfp}) of the partition function the specific heat of a damped free particle may be expressed as~\cite{Hanggi08NJP}
\begin{equation}
\begin{split}
C_S/k_B &= x^2_1 \psi'(x_1) + x^2_2 \psi'(x_2)\\
&\quad - \left (\frac{\hbar \beta \omega_D}{2 \pi} \right )^2 \psi'\left (\frac{\hbar \beta \omega_D}{2 \pi} \right ) -\frac{1}{2}\:,
\end{split}
\label{Cfp}
\end{equation}
where $\psi'(z)$ is the trigamma function~\cite{Abramowitz64Book}. This expression for the specific heat interpolates between the classical value $1/2$ (which is reached in the limit of an undamped particle $\gamma \to 0$ and in the high temperature limit $T \to \infty$) and the value $C=0$ for $T \to 0$ in accordance with the third law of thermodynamics.  For strong damping $\gamma/\omega_D >1$, the slope of the specific heat as a function of temperature at $T=0$ becomes negative with the consequence that a region of temperatures exists in which the specific heat is negative. This does not indicate any instability of the system but instead the circumstance that, with raising temperature,  more energy may be  stored in the interaction with the environment than flows  into the particle's kinetic energy. 

Using  Eq. (\ref{ZSfp}) for the partition function in combination with Eq. (\ref{S}) one finds an entropy that does not vanish for $T \to 0$. This seeming violation of the third law is due to the factorization of the partition function $Z_S$ in the classical free particle partition function and an environmental term. This approximation disregards the discreteness of the free particle spectrum, thereby leading to a violation of the third law.

An even more complicated behavior of the specific heat was reported by~\citet{Spreng13EPL} for other spectral densities than those leading to the Drude model. Coming from  positive values  at higher temperatures, the specific heat becomes negative for lower temperatures, then positive again for even lower temperatures, but formally fails to vanish at $T=0$. 
This apparent violation of the third law is again a consequence of a factorization of partition function analogous to Eq. (\ref{ZSfp}).
\subsubsection{Jaynes-Cummings-type model}\label{JCm}    
As another exactly solvable model we consider a two-level system interacting with an environment made up by a single harmonic oscillator. The Hamiltonian of the total system is
\begin{equation}
H_{\text{tot}} = \frac{\epsilon}{2} \sigma_z + \hbar \omega \left ( a^\dagger a + \frac{1}{2} \right ) + \kappa \sigma_z \left (a^\dagger a + \frac{1}{2} \right )\:.
\label{HJC}
\end{equation}
Here, $\sigma_z$ is the Pauli spin matrix and $a$ and $a^\dagger$ are annihilation and creation operators of the oscillator, respectively.   The parameters $\epsilon$, $\omega$, and $\kappa$ specify the energy difference of the bare two-level atom, the frequency of the oscillator, and the interaction strength between the two-level atom and the oscillator, respectively. 
Note that in the Hamiltonian (\ref{HJC}) the coupling term commutes with the first two terms describing the free evolution of the oscillator and spin, respectively. Hence, in contrast to the common Jaynes-Cummings model~\cite{Jaynes63IEEE}, the interaction is purely dephasing without causing transitions between the eigenstates of the isolated subsystems.
For the  spectrum of the total Hamiltonian, which is given by $E_{n,s} = \epsilon s/2+ (\hbar \omega + \kappa s)(n+1/2)$, $s= \pm 1$, $n =0,1,2,\dots$, to be  bounded from below, the inequality $\hbar \omega > |\kappa|$ must hold. The equality sign is excluded because  otherwise the spectrum contains a point with infinite degeneracy and therefore the system may not assume a canonical equilibrium.

The partition function of the total system is determined to yield~\cite{Campisi09JPA}
\begin{equation}
Z_{\text{tot}} = q_+ +q_-
\label{ZtotJC}
\end{equation}
with the abbreviations
\begin{equation}
q_\pm = \frac{e^{-\beta \hbar \omega/2} e^{\mp \beta( \epsilon + \kappa)}}{1- e^{-\beta (\hbar \omega \pm \kappa)}}\:.
\label{qpm}
\end{equation}
In combination with the partition function of the bare harmonic oscillator $Z_B=1/(2 \sinh \beta \hbar \omega/2)$, the partition function of the open two-level system becomes
\begin{equation}
Z_S= \left ( q_+ + q_- \right ) 2 \sinh \bb {\beta \hbar \omega/2 }\:,
\label{ZSJC}
\end{equation}
differing from the partition function of the bare two-level system $Z^0_S= 2 \cosh \big (\beta \epsilon /2 \big )$.

The Hamiltonian of mean force results in
\begin{equation}
H^*= \frac{\epsilon^*}{2} \sigma_z + \gamma 
\label{H8tl}
\end{equation}
with the renormalized level distance given by
\begin{equation}
\epsilon^* = \epsilon + \kappa +\frac{2}{\beta} \text{artanh} \left (\frac{e^{-\beta \hbar \omega} \sinh \beta \kappa}{1-e^{-\beta \hbar \omega} \cosh \beta \kappa} \right )\:,
\label{e*}
\end{equation}
and the energy shift $\gamma$ given by
\begin{equation}
\gamma= \frac{1}{2\beta} \ln \left (\frac{1- 2 e^{- \beta \hbar \omega} \cosh \beta \kappa + e^{-2 \beta \hbar \omega}}{\left (1- e^{-\beta \hbar \omega} \right )^2 } \right )\:.
\label{gamma}
\end{equation}
The change of the level-spacing $\Delta=\epsilon^*-\epsilon$ and the energy shift $\gamma$ vanishes for $\kappa=0$ and diverges when the absolute value of $\kappa$ approaches $\hbar \omega$, whereby $\Delta$ is an odd function and $\gamma$ is an even function of $\kappa$. 

\begin{figure}
\includegraphics[width=0.235\textwidth]{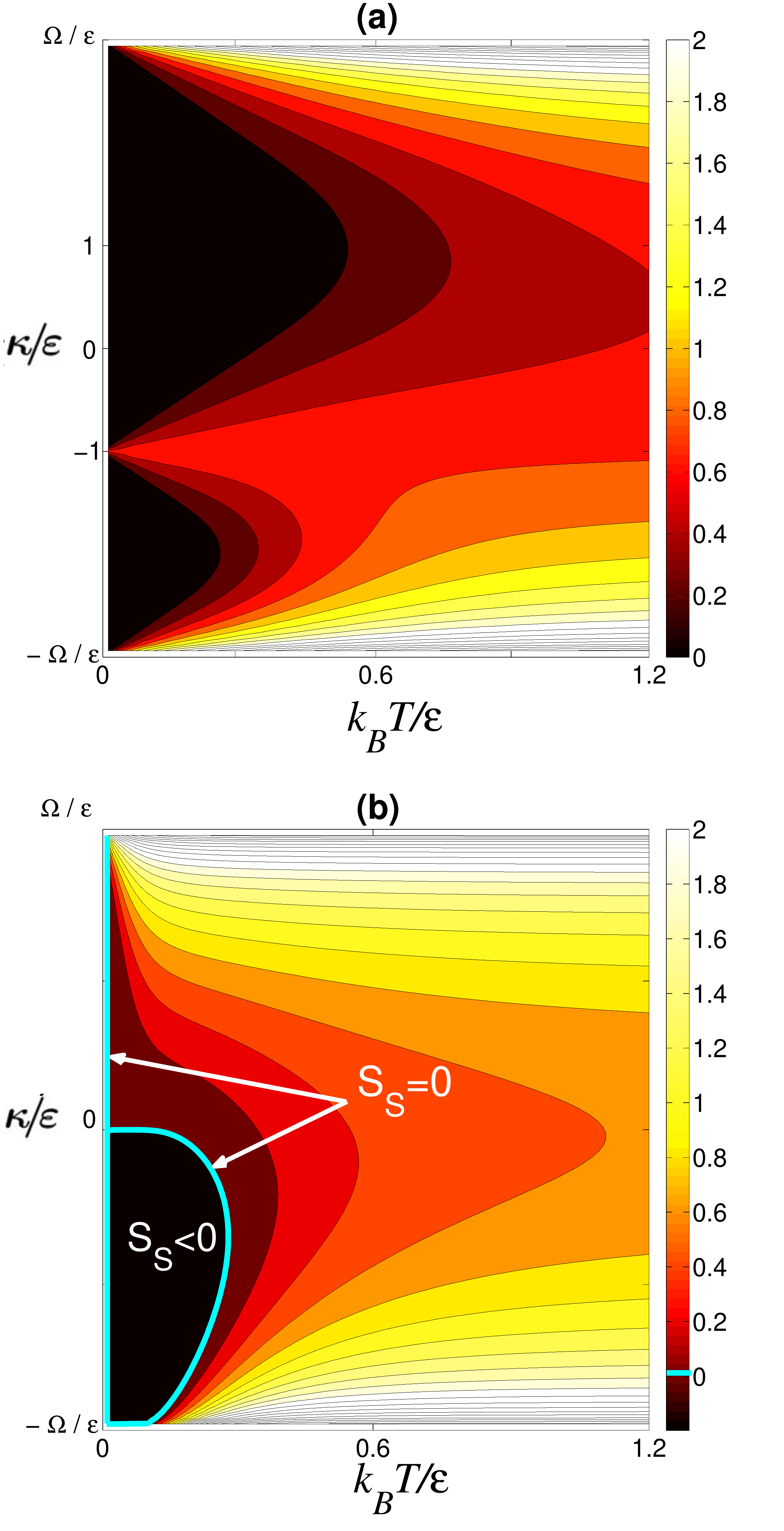}    
\includegraphics[width=0.235\textwidth]{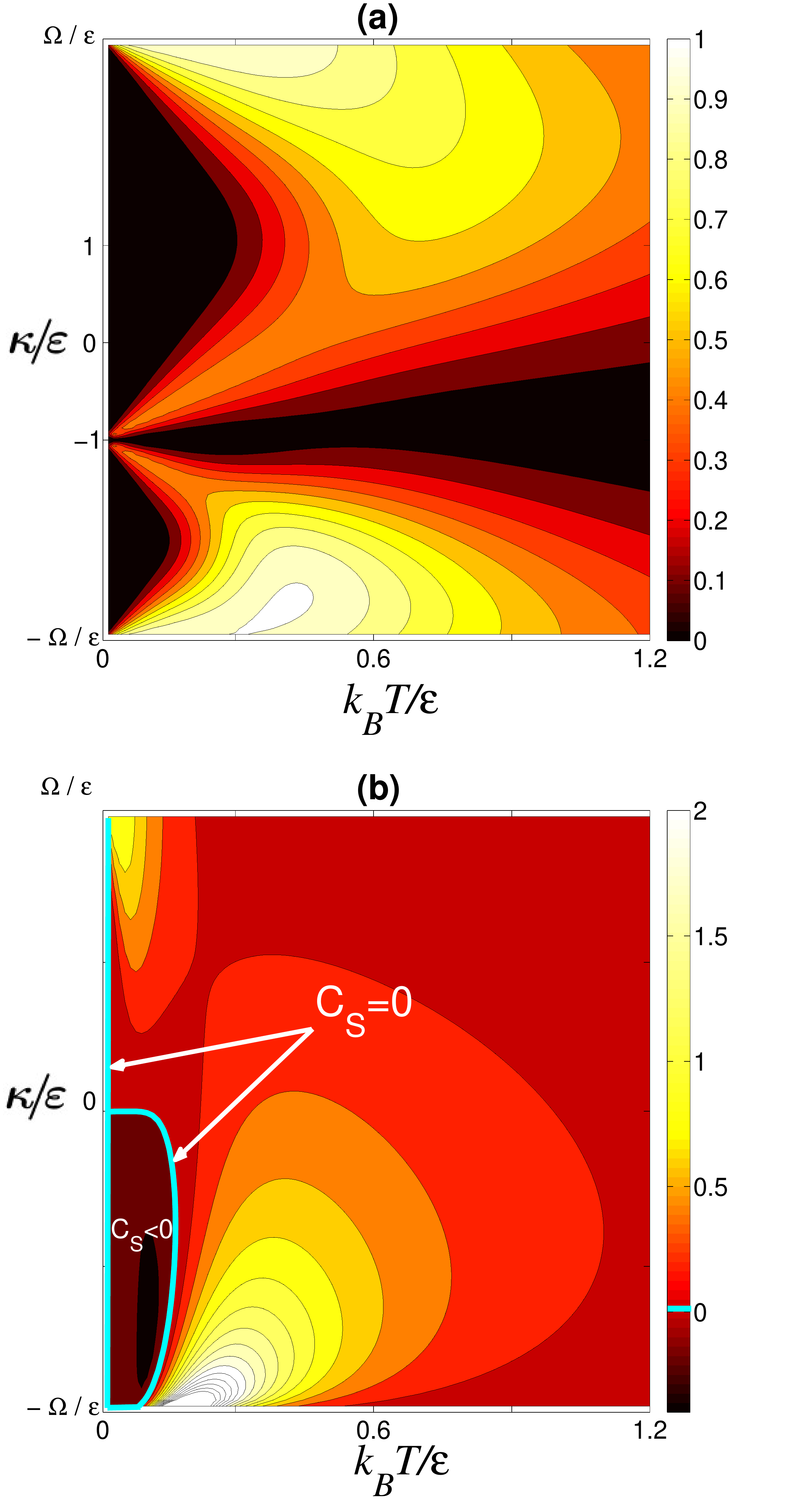}  
\caption{Role of interaction strength for entropy and specific heat of a two-level system with purely dephasing coupling to an oscillator [ Eq.\ref{HJC}]. The left panels display the entropy and the right panels the specific heat of a two-level atom in contact with a single harmonic oscillator as functions of the dimensionless temperature $k_B T/\epsilon$ varying on the horizontal axes and of the ratio of the coupling constant and the two-level atom splitting $\kappa/\epsilon$ within the allowed region $|\kappa| < \Omega \equiv \hbar \omega$. The ratio of the oscillator and the atom level splitting  is $\Omega/\epsilon =3$ in the upper two panels and $ \Omega/\epsilon =1/3$ in the lower two panels. From~\citet{Campisi09JPA}. }
\label{FigJCSC}
\end{figure}

The entropy $S_S = k_B\ln Z_S- k_B\beta \partial \ln Z_S/ \partial \beta$, given by  Eq. (\ref{SSH*}), and the specific heat $C_S= k_B \beta^2 \partial^2 \ln Z_s/ \partial \beta^2$  vanish in the limit $\beta \to \infty$, in agreement with the third law of thermodynamics. If the level-distance of the harmonic oscillator is less than that of the two-level atom, i.e., if $\hbar \omega < \epsilon$, both, the entropy and the specific heat become negative at low temperatures for negative coupling constants $\kappa <0$, see Fig.~\ref{FigJCSC}.
\subsubsection{Isotropic $XY$ spin chain}
\begin{figure}[h]
\includegraphics[width=0.4\textwidth]{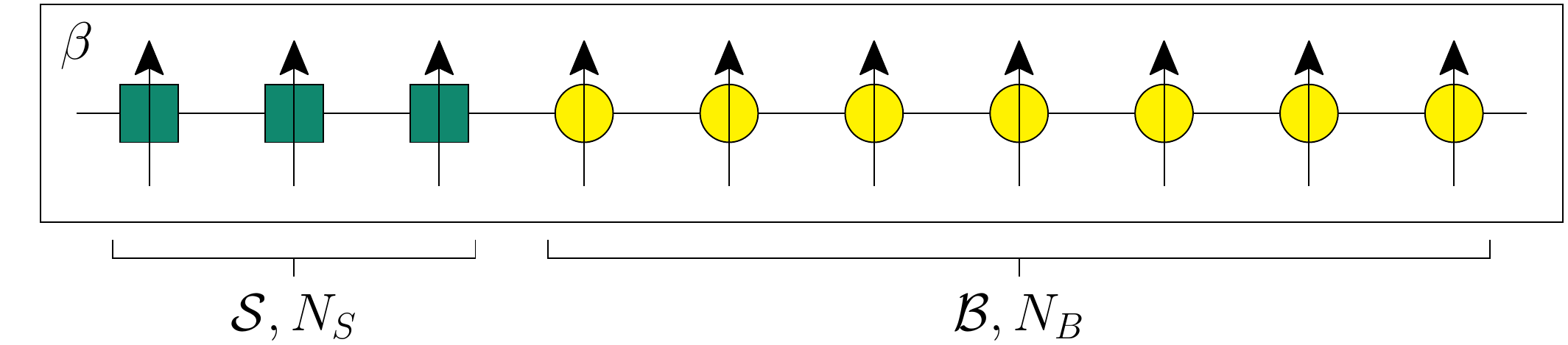}
\caption{$XY$-spin chain setup at inverse temperature $\beta$. The left $N_S$ spins depicted as squares constitute the system, while the remaining $N_B$ spins (circles) compose the environment. From~\citet{Campisi10CP}.}
\label{FigXY}
\end{figure}

\begin{figure}[h!]
\includegraphics[width=0.49\columnwidth]{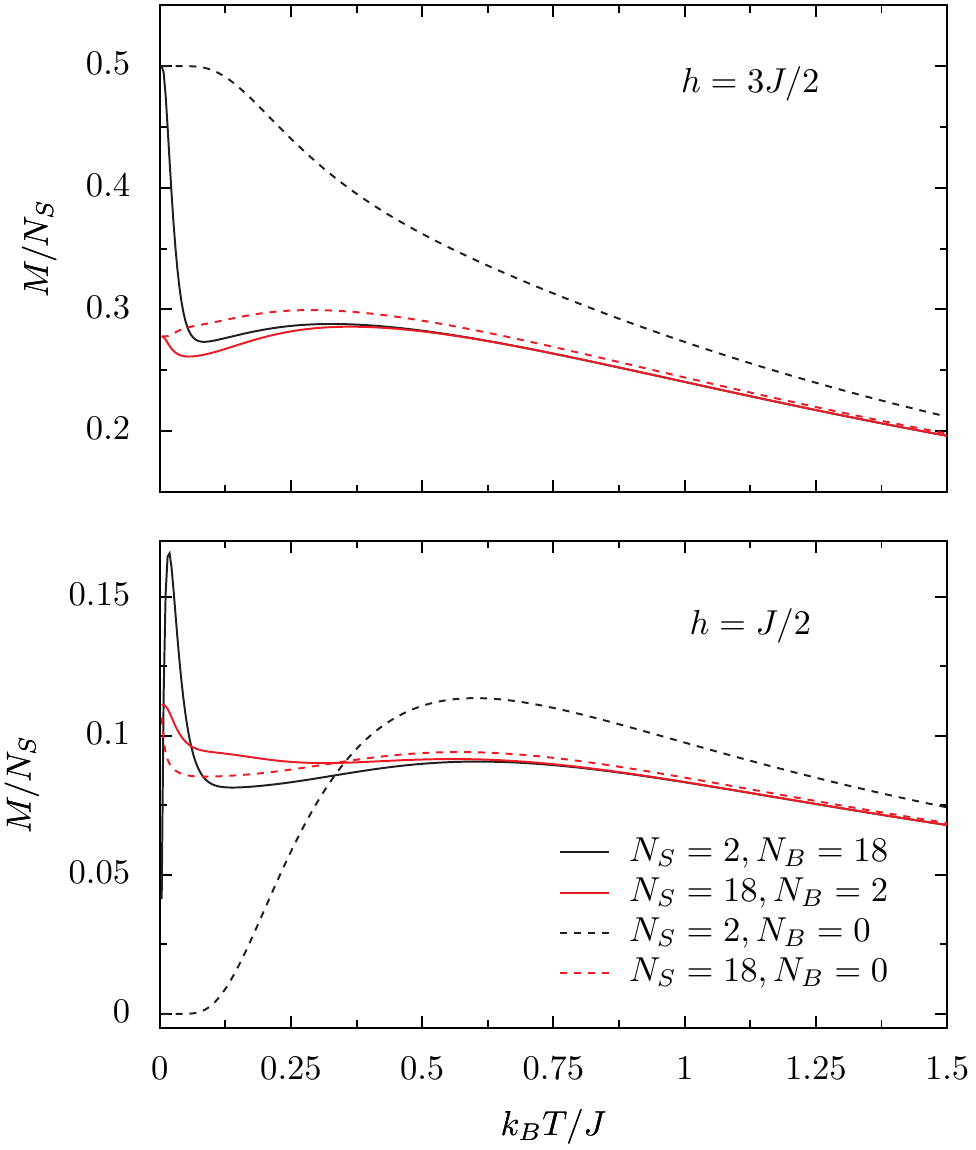}
\hfill
\includegraphics[width=0.49\columnwidth]{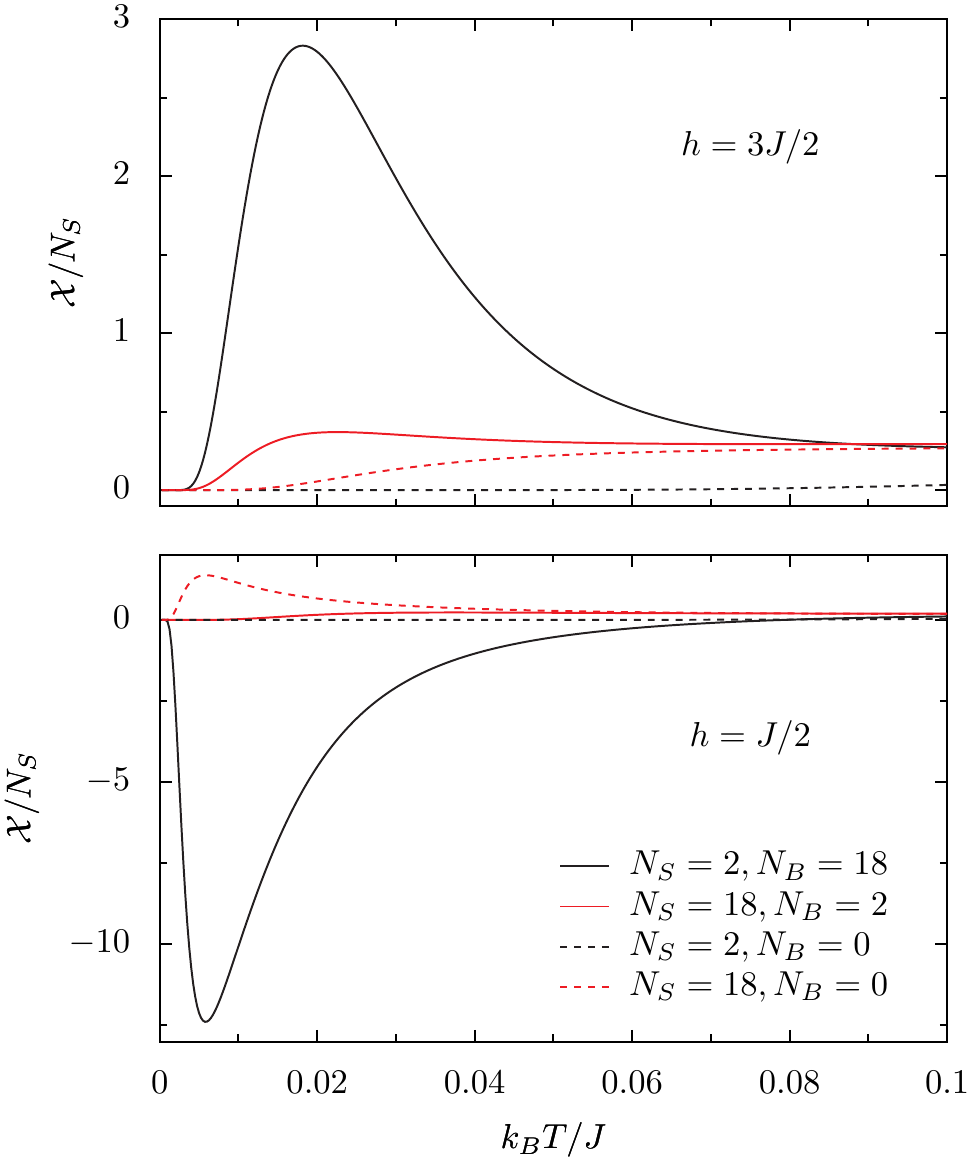}
\caption{ $XY$-spin chain: magnetization and susceptibility.The magnetization $M/N_S$ and the susceptibility $\chi/N_S$ per spin are displayed as functions of the reduced temperature for different chain lengths in the left and right panel, respectively. The upper and the lower panels refer to different on-site magnetic fields. From~\citet{Campisi10CP}.  }
\label{SpinMchi}
\end{figure}

We consider a linear chain of $N=N_S+N_B$ spins 1/2, of which the first $N_S$ spins constitute the system and the remaining $N_B$ spins constitute the environment sketched in Fig.~\ref{FigXY}. 
The chain has free ends and each spin experiences the same magnetic field $\mathsf{h}$ and nearest neighbor interaction of strength $J$. It is characterized by the Hamiltonian
\begin{equation}
H_N= \frac{\mathsf{h}}{2} \sum_{j=1}^N \sigma^z_j + \frac{J}{2} \sum_{j=1}^{N-1} \left ( \sigma^x_j\sigma^x_{j+1} + \sigma^y_j\sigma^y_{j+1} \right )\:.
\label{HXYN}
\end{equation}  
This model is exactly solvable by a Jordan-Wigner transformation~\cite{Mikeska77ZPB} yielding for the partition function
\begin{equation}
\begin{split}
Z_N &= \Tr e^{-\beta H_N}\\
&= e^{-\beta N \mathsf{h}/2} \prod_{k=1}^N \left ( 1+ e^{-\beta \lambda^{(N)}_k} \right )\:,
\end{split}
\label{ZXYN}
\end{equation}  
where
\begin{equation}
\lambda^{(N)}_k = \mathsf{h} - 2 J \cos\frac{\pi k}{N+1}\:.
\label{lNk}
\end{equation}
With the first $N_S$ spins of this chain as the system and the remaining $N_B=N-N_s$ spins as the bath, one can recast the total Hamiltonian in the form $H_{\text{tot}} \equiv H_N=H_S +H_B + H_{SB}$ with
\begin{align}
H_S &= \frac{\mathsf{h}}{2} \sum_{j=1}^{N_S} \sigma^z_j -\frac{J}{2} \sum_{j=1}^{N_S-1}\left ( \sigma^x_j \sigma^x_{j+1} + \sigma^y_j \sigma^y_{j+1} \right ),
\label{HSS}\\
H_B &= \frac{\mathsf{h}}{2} \sum_{j=N_S +1}^N \sigma^z_j -\frac{J}{2} \sum_{j=N_S+1}^{N-1}\left ( \sigma^x_j \sigma^x_{j+1} + \sigma^y_j \sigma^y_{j+1} \right ),
\label{HSB}\\
H_{SB}&= -\frac{J}{2} \left (\sigma^x_{N_S} \sigma^x_{N_S+1} + \sigma^y_{N_S} \sigma^y_{N_S+1} \right )\:. 
\label{HSSB}
\end{align}
The partition function of the system part follows as
\begin{equation}
\begin{split}
Z_S &=\frac{Z_N}{Z_{N-N_S}}\\
& = e^{-\beta N_S h/2} \frac{\prod_{k=1}^N \left (1+e^{-\beta \lambda^{(N)}_k} \right )}{ \prod_{k=N_S+1}^N \left (1+e^{-\beta \lambda^{(N-N_S)}_k} \right )}\:.
\end{split}
\label{ZSS}
\end{equation} 
\begin{figure}[h!]
\includegraphics[width=0.49\columnwidth]{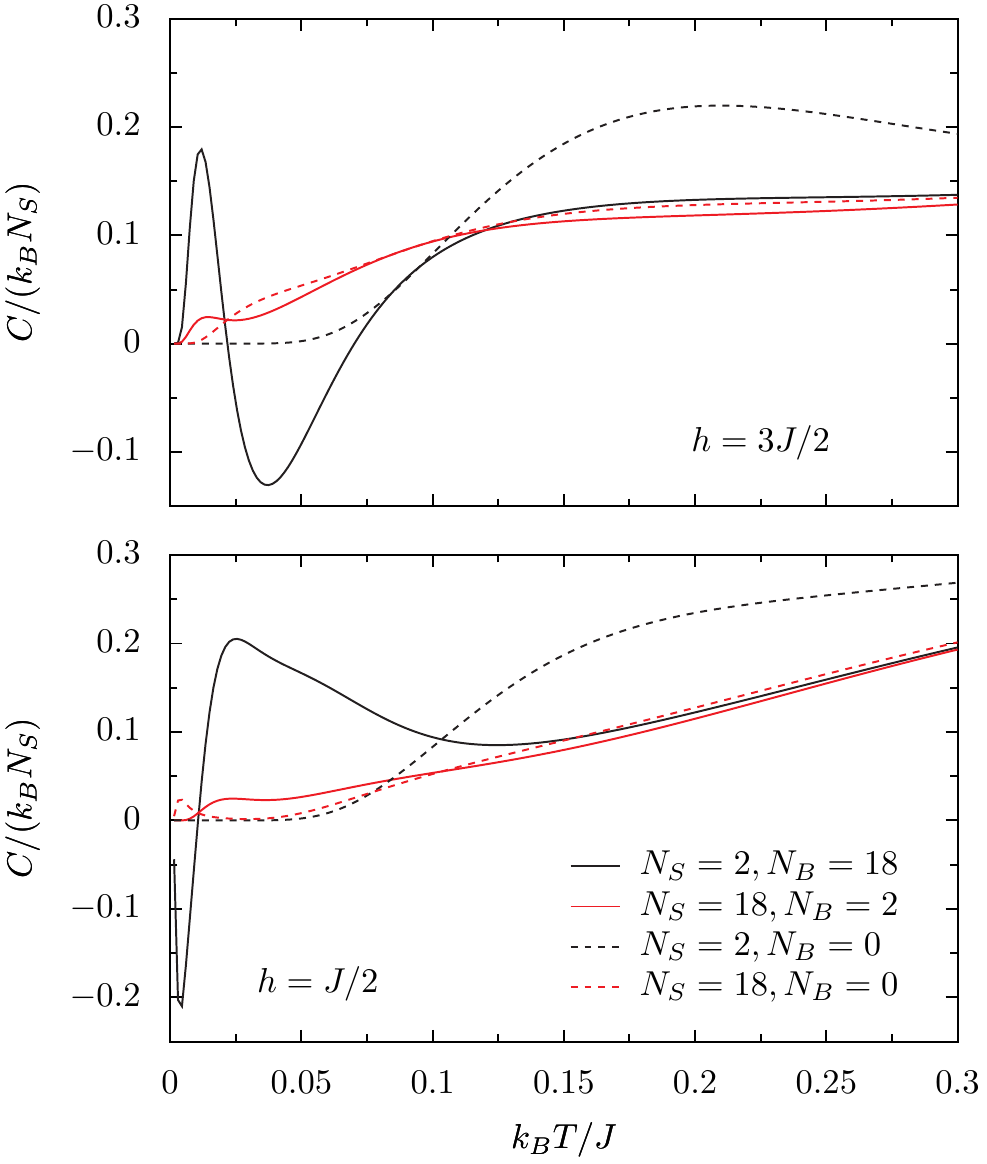}
\hfill
\includegraphics[width=0.49\columnwidth]{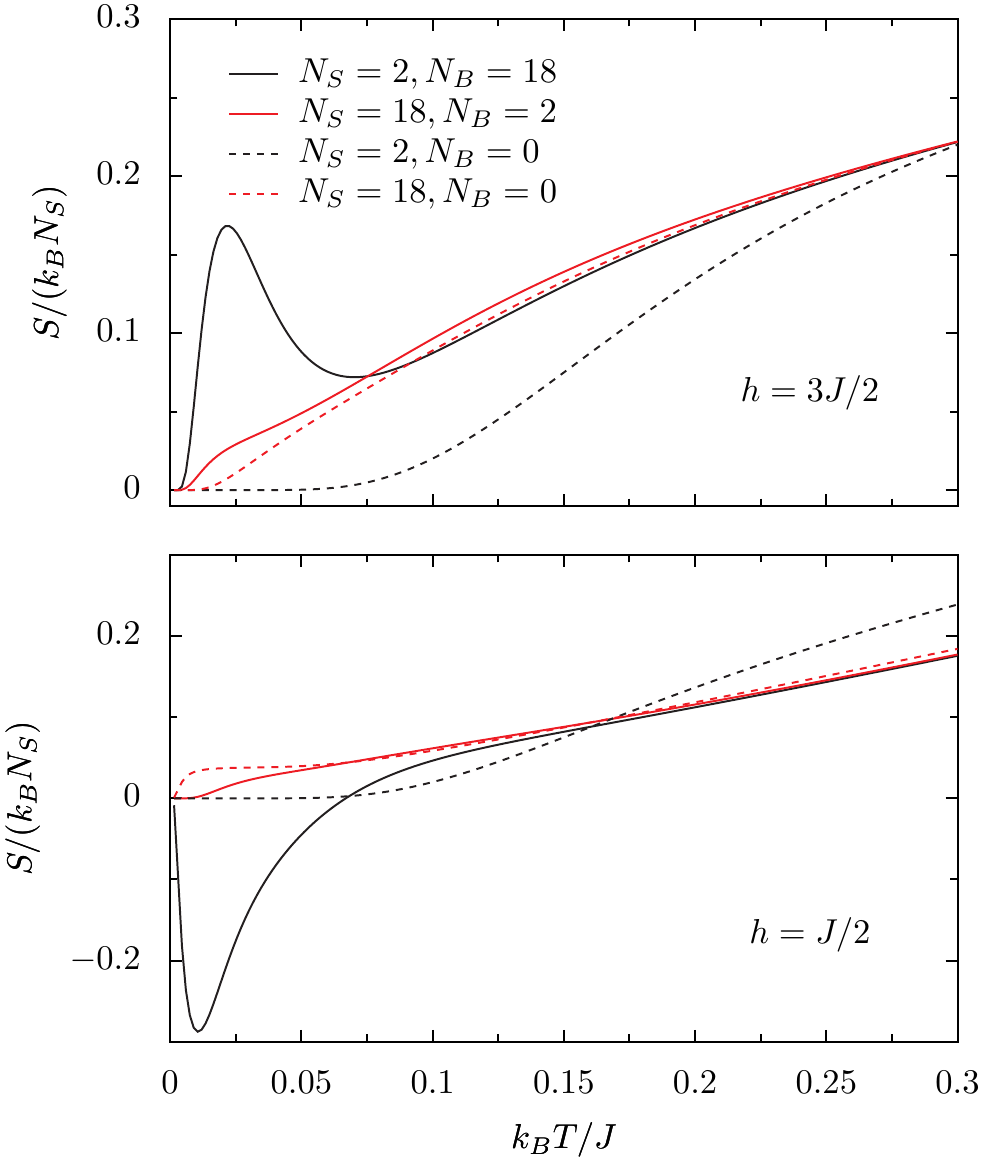}
\caption{$XY$-spin chain: specific heat and entropy. The specific heat and the entropy per spin of  spin chains are displayed in the left and right panels, respectively, for different magnetic fields $\mathsf{h}=3J/2$ (upper panels) and $\mathsf{h}=J/2$ (lower panels) and for different combinations of system and bath chain lengths as indicated by different lines as functions of the reduced temperature $k_B T/J$. From~\citet{Campisi10CP}.   }
\label{SpinSC}
\end{figure}
Figure~\ref{SpinSC} depicts the specific heat and entropy of open chains of different lengths for two different on-site magnetic fields $\mathsf{h}$ as functions of the temperature.  Depending on the strength of the on-site magnetic field and on the length of the open chain, the entropy and the specific heat display regions with negative values. As in the case of negative specific heat observed for a damped particle at low temperatures, this does not indicate any instability of the system but instead the capacity of the interaction with the environment to effectively store energy when the temperature is rising. 
For $T \to 0$, both the entropy and the specific heat vanish in accordance with the third law of thermodynamics.
The magnetization $M =\partial F_S/\partial \mathsf{h}|_\beta$ and the susceptibility $\chi =  \partial M/\partial \mathsf{h}|_\beta$ following with $F_S= - \beta^{-1} \ln Z_S$ from $Z_S$ according to  Eq.~(\ref{ZSS}) are illustrated for different on-site magnetic fields and chain lengths in Fig.~\ref{SpinMchi}.\footnote{Note that for the magnetization and the susceptibility the subtraction of the bare bath contribution is mandatory  because the magnetic field represents a {\it global} parameter.} For chains with large interaction $J>\sf{h}$, the susceptibility assumes negative values at low temperatures.  

\section{Going into nonequilibrium}\label{RMP_4}
There is a wide variety of circumstances that may drive a system out of thermal equilibrium, either temporarily or permanently~\cite{Keizer87Book,Zwanzig01Book}. Here we restrict ourselves to time-dependent changes of one or several of the system's parameters $\lambda$. 
The energy change involved in such a process is identified with the  work done on the system. To properly introduce this notion we first restrict ourselves to isolated systems,  then consider open systems in the absence of particle exchange. The pertinent Jarzynski equalities~\cite{Jarzynski97PRL} and Crooks relations~\cite{Crooks99PRE}  are reviewed in both situations.
\subsection{Work in thermally isolated systems}\label{Wiso}
\subsubsection{Classical systems}\label{Wisocl}
The work performed on an isolated system by a time-dependent variation of a system parameter is defined as the resulting change of the systems energy. For a classical system this prescription leads for the work $w$ performed by the variation of a system parameter $\lambda(t)$ between  $t=0$ and $t=\tau$ to  
\begin{equation}
w = H \bb {\mathbf{X}(\mathbf{x},\tau),\lambda(\tau) } - H\bb {\mathbf{x},\lambda(0) }\:,
\label{W}
\end{equation}   
where $\mathbf{x}$ denotes a point in the phase space of the system and $H \bb {\mathbf{x},\lambda(t) }$ denotes the system's Hamiltonian. It is assumed here that this time-dependent Hamiltonian is gauged such that its value coincides with the energy of the system~\footnote{\label{fn_gauge}For a more exhaustive discussion of the gauge-dependence of time-dependent Hamiltonians see~\citet{Campisi_Hanggi_Talkner11_aRMP}, cf. Sect. IIIA therein.}. With $\mathbf{X}(\mathbf{x},t)$ denoting the solution of Hamilton's equations of motion starting at $\mathbf{X}(\mathbf{x},0) =\mathbf{x}$ the difference of the Hamiltonians at the final point and at the initial point specifies the difference in energy and hence the work. Here we assume, following~\citet{Jarzynski97PRL}, that the work is determined by the difference of energies resulting from the full Hamiltonians. This work is known as inclusive, or Gibbsian, work. \citet{Bochkov81PAa,Bochkov81PAb} used a different definition of work. They assigned as final energy the value of the Hamiltonian at the initial parameter value at the propagated phase-space point, yielding
\begin{equation}
w^{\text{BK}} =   H \bb {\mathbf{X}(\mathbf{x},\tau),\lambda(0) } - H\bb {\mathbf{x},\lambda(0) }\:.
\label{WBK}
\end{equation}    
This work definition is also referred to as exclusive work~\cite{Jarzynski07CRP}. Here we restrict the discussion to the inclusive work defined in Eq.~(\ref{W}). More details on the exclusive work were given by \citet{Jarzynski07CRP,Campisi_Hanggi_Talkner11_aRMP}, and \citet{Campisi_Talkner11_PTRS}. 
Note that the value of the work depends on both the initial phase-space point $\mathbf{x}$ and the protocol $\Lambda =\{\lambda(t)|0 \leq t \leq \tau \}$, according to which the parameter $\lambda(t)$ changes in time. If the initial point $\mathbf{x}$ is randomly chosen, say, from the canonical equilibrium  pdf, the work becomes a random quantity that itself can be characterized by a pdf.

An equivalent formulation of the work defined in Eq. (\ref{W}) is obtained by rewriting the difference of the Hamiltonians as an integral of the time derivative over the duration of the protocol. Taking into account that the total time derivative of a Hamiltonian as it evolves along a trajectory coincides with its partial derivative one is lead to the following expression for the {\it classical work}:
\begin{equation}
w = \int_0^\tau dt \frac{\partial H \bb {\mathbf{X}(\mathbf{x},t),\lambda(t)}}{\partial \lambda(t)} \dot{\lambda}(t)\:,
\label{WP}
\end{equation}   
where $\dot{\lambda}(t)$ denotes the time derivative of $\lambda(t)$.\footnote{\label{fn_linearforce}For a forced parameter $\lambda(t)$ that couples to the system via a generalized coordinate $Q(\mathbf{x})$ according to $H\bb {\mathbf{x},\lambda(t) } = H_0(\mathbf{x}) - Q(\mathbf{x}) \lambda(t)$ the work expression simplifies to $w =-\int_0^\tau dt Q\bb {\mathbf{X}(\mathbf{x},t) } \dot{\lambda}(t)$. Hence, from an experimentally observed trajectory $Q\bb {\mathbf{X}(\mathbf{x},t) }$, $t \in [0,\tau]$ the work can be determined.}  
The value of the integrand at the time $t$ gives the power that is supplied to the system by the parameter variation at this instant.

For an initial phase-space pdf $\rho_0(\mathbf{x})$ the resulting pdf $p_\Lambda(w)$ of the work is given by~\cite{Campisi_Hanggi_Talkner11_aRMP}
\begin{equation}
\begin{split}
p_\Lambda(w) &= \int d \mathbf{x} \delta \left [ w\! -\! H\bb {\mathbf{X} (\mathbf{x},\tau), \lambda(\tau) } \!+\! H\bb {\mathbf{x},\lambda(0) } \right ] \rho_0(\mathbf{x})\\
&= \int d \mathbf{x} \delta \left ( w\! -\!\int_0^\tau d t \frac{ \partial H\bb {\mathbf{X} (\mathbf{x},t),\lambda(t) }}{\partial \lambda(t)} \dot{\lambda}(t) \right ) \rho_0(\mathbf{x} )\:.
\end{split}
\label{pLw}
\end{equation} 

Focusing on a {\it canonical} initial state at inverse temperature $\beta$, one obtains for the average of the exponentiated negative work per thermal energy $\beta w$ the Jarzynski equality~\cite{Jarzynski97PRL}  
\begin{equation}
\langle e^{-\beta w} \rangle_\Lambda = e^{- \beta\Delta F}\:,
\label{Ja}
\end{equation}  
where $\langle \cdot \rangle_\Lambda = \int dw \cdot p_\Lambda(w)$ denotes the average with respect to the work pdf (\ref{pLw}), and $\Delta F =F\bb {\beta,\lambda(\tau) } - F \bb {\beta, \lambda(0)}$ is the free energy difference of the canonical equilibrium states at the final and initial parameter values, both at the {\it same} temperature. For such isothermal processes $\Delta F$ corresponds to the maximal work that can be done reversibly by the system. The direct use of the Jarzynski equality, in both  numerical and real experiments, as a means of estimating the change of the free energy caused by a variation of a system's parameter is known to  often be severely hampered by statistical problems~\cite{Pohorille10JPCB,Deng17PRE,Lechner07JSM,Kim12PRE,Deng17Entropy}. For classical open systems in contact with a heat-bath that are described by a Langevin equation (see Sect.~\ref{open}), the work distribution in the presence of slow forcing can often be well approximated by a Gaussian distribution with a vanishing variance in the quasistatic limit~\cite{Hoppenau13JSM}. A similar result holds for open quantum Markovian processes described by a master equation of the Lindblad type~\cite{Miller19PRL}. As a consequence of the Jarzynski equation the irreversible work then vanishes as it does for a quasistatic isothermal macroscopic process. For more general system setups, both the absence and the presence   of finite work fluctuations in the quasistatic limit have been reported in the literature~\cite{Deng17PRE}.          

The most remarkable aspect of the Jarzynski equality is that it applies to an arbitrary protocol that is not restricted to be slow. 
In general, the finally reached state {\it differs} from the thermal equilibrium state corresponding to the final parameter value $\lambda(\tau)$. Note that for  thermally isolated forcing an equilibrium state that might be reached at large times will generally have a temperature differing from the initial one.  
Only if the system stays in weak contact with a thermal reservoir having the initial temperature, then the equilibrium state with the system free energy $F\bb { \beta,\lambda(\tau) }$ is approached for an infinitely slow protocol. For fast protocols an equilibration takes place  only after a sufficiently long time subsequent to the terminal protocol time $\tau$. During the equilibration the so-called irreversible work $w_{\text{irr}} =  w - \Delta F $ is taken by the reservoir.
With Jensen's inequality one obtains from the Jarzynski equality that the average of the irreversible work cannot become negative, in agreement with the second law of thermodynamics, i.e.,
\begin{equation}
\langle w \rangle_\Lambda \geq \Delta F\:
\label{2law}
\end{equation}
and consequently $\langle w_{\text{irr}} \rangle_\Lambda \geq 0$. The average irreversible work  is proportional to the Kulback-Leibler divergence of the actually reached final phase-space pdf $\rho(\tau)$ and the Gibbs state (\ref{rcan});
hence, $\langle w_{\text{irr}} \rangle_\Lambda= T S(\rho(\tau)||\rho(\beta,\lambda(\tau)))$~\cite{Kawai07PRL}.

To any forced process running according to a protocol $\Lambda=\{\lambda(t)|0\leq t \leq \tau \}$ a reverse force protocol $\bar{\Lambda}(t)=\{\epsilon_\lambda\lambda(\tau-t) |0 \leq t \leq  \tau \}$ can be assigned, where it is assumed that the instant Hamiltonians of the forward and backward processes are related by the time-reversal operation $H(\mathbf{q},\mathbf{p},\lambda) \to \bar{H}(\mathbf{q},\mathbf{p},\lambda) \equiv H(\mathbf{q},-\mathbf{p},\epsilon_\lambda\lambda)$ where $\lambda$ comprises all parameters on which the Hamiltonian depends, including those that remain fixed during the force protocol. Here $\epsilon_\lambda$ denotes the parity of the parameter $\lambda$ under time-reversal. For the pair of so-called forward and backward processes, both starting in a canonical equilibrium state at the same inverse temperature $\beta$ and at the respective parameter values $\lambda(0)$ and $\epsilon_\lambda \lambda(\tau)$,  the work pdfs $p_\Lambda(w)$ and $p_{\bar{\Lambda}}(w)$ are connected by the Crooks relation~\cite{Crooks99PRE} 
\begin{equation}
p_\Lambda(w) = e^{-\beta(\Delta F -w)} p_{\bar{\Lambda}}(-w)\:.
\label{Cr}
\end{equation}    
This implies that the occurrence of work smaller than the free energy change $\Delta F$ is exponentially small~\cite{Jarzynski11ARCMP}, i.e.
\begin{equation}
P\big [ w<\Delta F - \zeta \big ] \equiv \int_{-\infty}^{\Delta F - \zeta} dw p_\Lambda(w) \leq e^{-\beta \zeta}
\label{Pwzeta}
\end{equation}
This can be understood as a further specification of the second law of thermodynamics~\cite{Jarzynski07CRP,Jarzynski11ARCMP}. 

Multiplying both sides of the Crooks relation~(\ref{Cr}) by the factor $e^{-\beta w}$ and integrating over the work one recovers the Jarzynski equality (\ref{Ja}).

\subsubsection{Quantum systems: The two-point-projective-energy-measurement scheme (TPPEMS)}\label{Wisoqm}
At first glance the translation of a classical work expression to quantum mechanics might seem obvious by simply replacing the Hamiltonian functions in the equivalent classical work definitions [Eqs.~(\ref{W}) and (\ref{WP})] with the corresponding operators. This naive approach fails to lead to  a proper work operator for various reasons. While in the classical expression (\ref{W}) the Hamiltonians are evaluated at specific phase-space points that are connected by a trajectory of the Hamiltonian dynamics, the corresponding Hamiltonian operators are independent of the initial and time-evolved state of the system. For the equivalent classical work expression (\ref{WP}) the phase-space trajectory connecting these states has to be known. Owing to the lack of classical trajectories in the quantum case, however, this expression cannot be directly converted to a work expression for the quantum world.~\footnote{\label{Bohm}The work calculated along Bohmian trajectories turns out to depend on the particular representation of the initial state in terms of pure states~\cite{Sampaio18PRA} and hence cannot be considered a measurable quantity.} Further, in classical systems the specification of the energy can, in principle, be performed without any perturbation of the system. For a quantum system, gaining information about the system, from which its energy can be inferred, necessitates the interaction with an auxiliary system such as a measurement apparatus which, in turn, causes a {\it back-action} on the considered system. Hence, the specific tools employed to identify energies of a system from which the work is determined constitute a relevant part of any operational definition of quantum work.

For the sake of definiteness, we consider the TPPEMS. In this measurement scheme a projective energy measurement is performed on the system in the state $\rho_0$ immediately before the force protocol $\Lambda$ starts, i.e., according to our previous convention, at $t=0$, and the second one at $t=\tau$ immediately after the  protocol is finished~\cite{Kurchan00arX,Tasaki00arX,Talkner07PRE}.  The joint probability to observe the energies $E_n(0)$ and $E_m(\tau)$ is then given by
\begin{equation}
p_\Lambda(m,n) = \Tr P_m(\tau) U_\Lambda P_n(0) \rho_0 P_n(0) U^\dagger_\Lambda\:.
\label{pLmn}             
\end{equation}
Here $P_k(t)$ denotes the operator projecting onto the eigenspace of the Hamiltonian $H\bb { \lambda(t) } = \sum_k E_k(t) P_k(t)$ with eigenenergy $E_k(t)$.\footnote{\label{fn_spectra}In this general setting Hamiltonians with spectra containing accumulation points and continuous parts can also be considered. Because of the finite resolution of any measurement apparatus the probability $p_\Lambda(m,n)$ must then be replaced by the probability $p_\Lambda(A,B) =\Tr P_A(\tau) U_\Lambda P_B(0) \rho_0 P_B(0) U^\dagger_\Lambda $ where $P_C(t)$ projects on all eigen-states with energies $E(t) \in C$, $C=A,B$ being subsets of the spectrum captured by the two measurements at $t=0,\tau$.} Because of the first measurement the initial density matrix $\rho_0$ is projected onto the subspace with $E_n(0)$ and subsequently propagated by the unitary time-evolution operator 
\begin{equation}
U_\Lambda = \mathcal{T} e^{-i \int_0^\tau dt H\bb {\lambda(t) }/\hbar}\:,
\label{U0tau}
\end{equation}
 where   $\mathcal{T}$ denotes the chronological time-ordering operator. The work pdf $p_\Lambda(w)$ follows from Eq. (\ref{pLmn}) as 
\begin{equation}
p_\Lambda(w) = \sum_{m,n} \delta \bb {w - E_m(\tau) + E_n(0) } p_\Lambda(m,n)\:.
\label{pLwQ}
\end{equation}       
Equivalently, the work statistics can be described by the characteristic function $G_\Lambda(u) = \int dw e^{i u w} p_\Lambda(w)$, which assumes the form~\cite{Talkner07PRE}
\begin{equation}
G_\Lambda(u) = \Tr \Big (e^{i u H^H \bb {\lambda(\tau) }}  e^{-i u H \bb { \lambda(0) }} \bar{\rho}_0 \Big )
\label{GLu}
\end{equation}
with the Hamiltonian $H^H \bb {\lambda(\tau) } = U^\dagger_\Lambda H \bb {\lambda(\tau) } U_\Lambda$ in the Heisenberg picture; further, $\bar{\rho}_0 = \sum_n P_n(0) \rho_0 P_n(0)$ is the initial density matrix projected onto the energy basis~\cite{Talkner08PRE}. Here it is worth noting that the projection of the initial state due to the first energy measurement has an impact on the average work $\langle w \rangle_\Lambda = \Tr \left [H^H\bb {\lambda(\tau) } - H\bb {\lambda(0) } \right ] \bar{\rho}_0$, which differs from the difference of the average energies at the end and the beginning of the force protocol. This average energy difference  is given by $\Delta \langle E \rangle  = \Tr \left [H^H\bb {\lambda(\tau) } - H\bb {\lambda(0) } \right ] \rho_0 $ and is also  known as the untouched work~\cite{Talkner16aPRE}. The difference between  $\langle w \rangle_\Lambda$ and $\Delta \langle E \rangle$ vanishes only if the initial density matrix is diagonal with respect to the energy basis of the initial Hamiltonian. In general, $\Delta \langle E \rangle - \langle w \rangle_\Lambda$ may be either positive or negative, and hence energy may seemingly be gained or lost in the TPPEMS if compared to the change of the average energies.\footnote{\label{fn_twolevel}As a simple example, one may consider a two-level atom whose initial density matrix has diagonal elements  $p$ and $1-p$ and nondiagonal element $q$ and $q^*$ (the asterix indicates  complex conjugation) with $p (1-p) \geq |q|^2$ when specified in the  energy eigenbasis of the initial Hamiltonian $H_0$. The force protocol consists of a sudden quench of the Hamiltonian with diagonal elements $h_1$ and $h_2$ and nondiagonal elements $c$ and $c^*$, again with respect to the eigenbasis of $H_0$. The energy mismatch then becomes $\Delta \langle E \rangle - \langle w \rangle = c q^* + c^* q$, an expression that can take either sign.} 
Attempts to interpret the energy mismatch in the spirit of Landauer's principle \cite{LandauerIBM61} as  equivalent to the gain of information proposed e.g. by \citet{KammerlanderSR16} and \citet{DeffnerPRE16}, however, do not explain why the first energy measurement is energetically relevant but not the second one. Moreover, to translate information that can be quantified as negative Shannon entropy one needs to make contact with a thermal bath, even though the system is isolated during the entire force protocol. In particular, when the mismatch has a finite value because of a nonthermal  initial state  there is no natural choice to assign a temperature value and the information gain has no obvious energy equivalent.       

Note that the characteristic function of work differs in form from one
that specifies the statistics of an observable $\mathcal{O}$. In the latter case it had to take the form $G(u) =\Tr e^{i u \mathcal{O}} \rho$. Hence, one cannot characterize work by an observable~\cite{Talkner07PRE}. Yet the characteristic function of work (\ref{GLu}) satisfies the formal sufficient 
and necessary conditions of being the Fourier transform of a probability density. These are (i) $G_\Lambda(0)=1$, (ii) $|G_\Lambda(u)|\leq 1$ and (iii) $\int du dv f^*(u) G_\Lambda(u-v) f(v) \geq 0$ for all integrable complex valued functions $f(u)$~\cite{Lukacs70Book}.\footnote{Condition (i) follows immediately with $\Tr \rho_o =1$, condition (ii) is the consequence of $|\Tr A \rho | \leq || A||_B || \rho_0 ||_{TC}$ where $||A||_B=1$ is the operator norm of the unitary operator $A = e^{iu H^H\bb {\lambda(\tau) } } e^{-iu H \big ( \lambda(0) \big ) }$ and $||\rho_0||_{TC} =1$ the trace-class norm of the initial density matrix; for the definitions of the different operator norms see~\cite{Schatten50Book}. Finally, with  $C = \int du f(u) e^{-i u H^H\bb { \lambda(\tau) } } e^{iu H \bb {\lambda(0) }}$ condition (iii) follows according to  $\int du dv f^*(u) G_\Lambda(u-v) f(v) = \Tr C^\dagger C \bar{\rho}_o \geq  0$.}   

\citet{PerarnauPRL17} demonstrated that no measurement scheme\footnote{\label{fn_measurement}We refer to a  measurement scheme as a family of completely positive maps specifying the states after a selective measurement together with the probabilities of finding all possible results; for more details see, e.g., Chap. 2.4 of \citet{Breuer02Book} } of work exists that is linear in the initial state of the system and for which the following two conditions are simultaneously satisfied: (1) The average work agrees with the difference of the average final and initial energies for any initial state, and (2) the resulting work statistics agrees for diagonal initial states (i.e., $\bar{\rho}_0 = \rho_0$) with the TPPEMS result [Eq.~\ref{pLwQ}]. A series of alternative attempts to define work in quantum systems other than by the TPPEMS were analyzed in view of this no-go theorem by~\citet{Baumer18}. We note that there exist two point measurement schemes using generalized energy measurements~\cite{Watanabe14PRE} as well as generalized work measurements~\cite{Talkner16aPRE} for which it is possible to reconstruct the work distribution of the TPPEMS.

For systems initially in a Gibbs state, $\rho_0 = \rho\bb {\beta, \lambda(0) } =Z^{-1}\bb {\beta, \lambda(0) } e^{-\beta H\bb {\lambda(0) }}$
the TPPEMS leads to the quantum Jarzynski equality [Eq.~\ref{Ja}]. Likewise, the average of the irreversible work can be written as the Kullback-Leibler divergence between the actual final state and the Gibbs state at the initial temperature and final parameter values~\cite{Deffner10PRL}. Further, the average irreversible work $\langle w_{\text{irr}} \rangle_\Lambda$ can be subdivided into a part that is due to coherences with respect to the final energy eigen-basis and another part that is caused by deviations of the finally reached populations of the final energy states from those of a canonical distribution with the final Hamiltonian at the initial temperature~\cite{Francica19PRE}.  
Moreover, for Hamiltonians transforming under time reversal as $\bar{H}(\lambda) \equiv \theta H(\lambda) \theta^\dagger = H(\epsilon_\lambda \lambda)$,  the Crooks relation (\ref{Cr}) is obeyed in exactly the same way as in classical systems~\cite{Talkner07JPA,Tasaki00arX}. Here $\theta$ denotes the antiunitary time-reversal operator \cite{Messiah62Book} and $\epsilon_{\lambda}$ is the parity of the parameter $\lambda$ under time reversal.      
In general, for initial states differing from Gibbs states no fluctuation relations exist. Exceptions are a microcanonical initial state for which a Crooks type relation holds yet a Jarzynski equality is not known~\cite{Talkner08PRE,Talkner13NJP}. For  grand canonical initial states both types of fluctuation relations hold. These relations  involve both work and exchanged particle numbers together with the difference of the respective grand potentials~\cite{Yi12PRE}. 

We note that it is not possible  to mutate the classical expression (\ref{WP}) into a quantum-mechanical form that is compatible with the fluctuation relations of Crooks and Jarzynski. A projective measurement of the  {\it work operator} $W = \int_0^\tau dt \mathcal{P}(t)$ defined in terms of a {\it power operator} $\mathcal{P}(t) = \dot{\lambda}(t) \partial H^H\bb {\lambda(t) }/\partial \lambda(t) $ yields on average the difference of the energy averages at the final and the initial times and, therefore, according to the findings of \citet{PerarnauPRL17},  cannot yield the {\it work statistics} of the TPPEMS for an initial state that is diagonal in the energy basis. Even the weaker requirement of satisfying the Jarzynski equality is not fulfilled \cite{Engel07EPL}. In addition, a continuous weak measurement of the power operator $\mathcal{P}(t)$ turns out to be  incompatible with the fluctuation theorems \cite{Venkatesh15NJP}. 

Finally, we remark that for a two point generalized energy measurement scheme the requirement that the Crooks relation is satisfied  already restricts  the allowed types of measurements to projective ones for systems with an infinite-dimensional Hilbert space. For systems with finite-dimensional Hilbert spaces slightly more general measurements are possible; the measurements still need to be error free, meaning that if the state in which the system is measured is an eigenstate belonging to a particular energy value, this energy value must be detected with certainty. For further details and  the restrictions imposed by the Jarzynski equality we refer to the literature \cite{Ito19PRA,Watanabe14aPRE}.
\subsection{Work in open systems} \label{open} 
The work applied to an open system, which is part of a large closed system described by a Hamiltonian, as specified in Eq. (\ref{Htot}), agrees with the work done on the total system only if system parameters are changed  that  influence neither the interaction nor the bath Hamiltonian. If the latter condition is not fulfilled,  the work done on the total system can be identified with the change of the total system but the work done on the open system cannot be defined.\footnote{\label{fn_global}If, for example,  both the system and the environment are electrically polarizable, the change of an externally controlled electromagnetic field directly affects both constituents.}

We start with a discussion of work in open quantum systems and later specialize to the respective classical case.
\subsubsection{Work in open quantum systems}\label{oqs}
The statistics of work performed on an open system upon changing a system parameter $\lambda(t)$ according to a specified protocol $\Lambda$ is formally determined by the same expression (\ref{pLwQ}) as in the case of a closed system where all quantities refer to the total system~\cite{Campisi09PRL}. Specifically, in  Eq.~(\ref{pLwQ}), $E_m(t)$  
indicates the eigenvalue 
of the total system Hamiltonian $H_{\text{tot}}\bb {\lambda(t) } = H_S\bb {\lambda(t) }  + H_B + H_{SB}$. Likewise, the time-evolution operator 
\begin{equation}
U_\Lambda = \mathcal{T} e^{i \int_0^\tau dt H_{\text{tot}}\bb {\lambda (t) }/\hbar} 
\label{Ula}
\end{equation}
in Eq.~(\ref{GLu}) is governed by the total Hamiltonian, and the density matrix $\rho_0$ specifies the total initial state.  
For a canonical initial state of the total system, $e^{-\beta H_{\text{tot}}\bb {\lambda(0)}}/ \Tr e^{-\beta H_{\text{tot}}\bb {\lambda(0)}}$  the fluctuation relations of Crooks and Jarzynski follow, with the free energy difference $\Delta F = \Delta F_S$ holding because of $F_S(\beta,\lambda) = F_{\text{tot}}(\beta,\lambda) -F_B(\beta)$ [ see Eq. (\ref{XiXi})] and the fact that the bare bath free energy is independent of the system parameter $\lambda$.\footnote{\label{fn_timerev}We note that the backward protocol also requires the reversal of all parameters transforming oddly under time reversal like magnetic fields, even if they are kept constant during the protocol or affect only the bath dynamics.}  Hence, one has
\begin{align}
p_\Lambda(w) &= e^{-\beta(\Delta F_S -w)} p_{\bar{\Lambda}}(-w)\:,
\label{Crops}\\
\langle e^{-\beta w} \rangle_\Lambda &= e^{-\beta \Delta F_S}\:,
\label{Jop}
\end{align}
i.e. the fluctuation relations continue to hold for open quantum systems that start in a total canonical equilibrium state independently of  the coupling strength between system and bath and also {\it irrespective} of the nature of the open system's dynamics, with work and free energy differences both relating to the open system~\cite{Campisi09PRL}.  This fact has raised doubts as to  whether a work statistics within the TPPEMS contains any quantum aspects at all. These doubts have been removed by several case studies~\cite{Talkner08bPRE,Yi11PRE,Yi12PRE,Deffner08PRE}  
and also by
the identification of quantum  coherences generated during a force protocol~\cite{Francica19PRE,Blattmann17PRA,Miller18E} and by the investigation of quantum mechanically generated deviations of the work statistics from their classical Gaussian form for almost quasistatic isothermal processes~\cite{Baumer19Q}.

Yet, in an experiment projective measurements of the total system Hamiltonian are to be imposed. Not only that they are difficult to perform, but the generally small difference between the much larger energies of the final and the initial state of the total system must be considered a severe practical limitation of the TPPEMS.     
\subsubsection{Work in open classical systems} \label{ocs}
The arguments leading to the quantum fluctuation relations for open systems can be repeated almost literally for classical open systems~\cite{Jarzynski04JSM}.  The dynamics of the classical total system is governed by a Hamiltonian of the form
\begin{equation}
H_{\text{tot}}(\mathbf{z}, \lambda(t)) = H_S(\mathbf{x}, \lambda) + H_B(\mathbf{y}) +H_{SB}(\mathbf{z})\:,
\label{Htotc}
\end{equation}
where $\mathbf{z} = (\mathbf{x}, \mathbf{y} )$ indicates a point in the phase-space of the total system with components $\mathbf{x}$ and $\mathbf{y}$ specifying  phase space points of the system and its environment, respectively.
Again, only the system Hamiltonian $H(\mathbf{x},\lambda)$ depends on the parameters $\lambda$ that are subject to the protocol $\Lambda$. In analogy to Eqs. (\ref{W}) and (\ref{WP}) the work can be expressed either as the energy difference of the total system, or as an integral of the power, to yield
\begin{align}
w &= H_{\text{tot}} \bb {\mathbf{Z}(\mathbf{z},\tau),\lambda(\tau) } - H_{\text{tot}} \bb {\mathbf{z},\lambda(0) } \label{wHtot}\\
&= \int_0^\tau dt \frac{\partial H_s \bb {\mathbf{X}(\mathbf{z},t),\lambda(t) }}{\lambda(t)} \dot{\lambda}(t)\:,
\label{wP}
\end{align} 
where $\mathbf{Z}(\mathbf{z},t)$ denotes the trajectory in the full phase space starting at $\mathbf{z}$ and $\mathbf{X}(\mathbf{z},t)$ is the projection of $\mathbf{Z}(\mathbf{z},t)$ onto the phase space of the open system.  Therefore, for classical open systems measuring the total energy can be circumvented. 
Instead, the system trajectories during the protocol have to be monitored and used to calculate the supplied power. Hence, as for closed systems, based on the power supplied during the protocol, the work  done on an open system can be determined from the sole observation of the system trajectories.

In correspondence with Eq. (\ref{pLw}), the work pdf becomes
\begin{equation}
\begin{split}
p_\Lambda(w) &=  \int d \mathbf{z}\: \delta \left ( w- \int_0^\tau dt \frac{\partial H_S \bb {\mathbf{X}(\mathbf{z},t),\lambda(t) }}{\partial \lambda(t)} \dot{\lambda}(t) \right )\\ 
&\quad \times\rho_0(\mathbf{z})    
\end{split}
\label{pLcl}
\end{equation}
with the total phase-space pdf $\rho_0(\mathbf{z})$ characterizing the initial state. For total systems initially staying in a canonical state, the Jarzynski and Crooks relations follow~\cite{Jarzynski04JSM}. As in the quantum case, the fluctuation relations hold for open classical systems irrespective of the kind of stochastic dynamics of the open system. For Markovian processes the fluctuation relations are derived directly from Fokker-Planck equations~\cite{Kurchan98JPA,Hummer01PNAS,Hatano01PRL} and master equations~\cite{Jarzynski97PRE,Gaspard04JCP,Harris07JSM,Esposito09RMP}.   
\section{Fluctuating thermodynamic potentials}\label{RMP_5}
In this section we first restrict ourselves to a discussion of fluctuating thermodynamic potentials in thermal equilibrium for  classical systems and only later comment on quantum mechanics. 
Inspired by the fact that the work performed on a system is a fluctuating quantity, one may ask whether it would not also be possible and even meaningful to consider fluctuating heat. Assuming the validity of an instantaneous first law one may construct from the fluctuating work and heat a likewise fluctuating internal energy
as proposed in stochastic energetics~\cite{Sekimoto98PTP,Sekimoto10Book}. Additional fluctuating thermodynamic potentials, in particular, fluctuating entropy, are considered in stochastic thermodynamics~\cite{Seifert05PRL,Seifert12RPP,vandenBroeck15PhysA}.
\subsection{Fluctuating internal energy}\label{Fie}
For a classical system in contact with its environment a fluctuating internal energy, as it is postulated using stochastic energetics~\cite{Sekimoto10Book} and stochastic thermodynamics~\cite{Seifert12RPP}, is supposed to assign to each momentary state $\mathbf{x}$ of the open system a uniquely defined energy value. In general, one might  expect such an assignment  to also require some information about the actual state of the environment.\footnote{\label{fn_fie}The dependence of the energy of an open system on the instantaneous state of the environment can be illustrated by the example of a dipolar molecule in a polar fluid. The magnitude and orientation of the molecule's electrical dipole moment relative to the local electric field determine a contribution to the energy of the molecule. Because the state of the fluid is not static in thermal equilibrium, and hence the orientation and magnitude of the local electrical field produced by the fluid surrounding the molecule fluctuate, an environmentally state-dependent, and therefore random, contribution  to the molecule's energy results.} In the sequel we therefore consider the more general {\it hypothesis} that a fluctuating internal energy can be characterized by a function $e(\mathbf{z},\beta,\lambda)$, where, as introduced in  
Sec.~\ref{ocs}, 
points in the phase space of the open system are denoted by $\mathbf{z}=(\mathbf{y},\mathbf{x})$ with an environmental component $\mathbf{y}$ and an open system component $\mathbf{x}$.   
Because the microstate $\mathbf{y}$ of the environment is not be monitored, the hypothetical fluctuating internal  energy $e(\mathbf{z}, \beta,\lambda)$ has to be considered as random field, where the random variable $\mathbf{y}$ is distributed according to the  conditional pdf $w(\mathbf{y}|\mathbf{x})$ defined in Eq.~(\ref{wyx}). 
A basic requirement for a fluctuating internal energy is that its average  with respect to the canonical equilibrium state of the total system must coincide with the internal energy $U_S$ of the total system, so that with $\rho_{\text{tot}}(\mathbf{z},\beta,\lambda)$ in Eq. (\ref{rtot})
\begin{equation}
U_S(\beta,\lambda)= \int d \mathbf{z} e(\mathbf{z}, \beta,\lambda) \rho_{\text{tot}} (\mathbf{z},\beta, \lambda)\:.
\label{Ue}
\end{equation}
Combined with  Eq.~(\ref{USH*}) for the internal energy one finds that any fluctuating internal energy must be of the form
\begin{equation}
\begin{split}
e(\mathbf{z},\beta,\lambda) &= \frac{\partial}{\partial \beta} [\beta H^*(\mathbf{x},\beta,\lambda)] + h_u(\mathbf{z},\beta,\lambda)\\
&= \langle H_{\text{tot}}|\mathbf{x} \rangle - \langle H_B \rangle_B + h_u(\mathbf{z},\beta,\lambda)\:,
\end{split}
\label{ieH*}
\end{equation}
where $h_u(\mathbf{z},\beta,\lambda) \in \mathcal{N}_{\beta,\lambda}$ is a random field with vanishing mean value in thermal equilibrium. Accordingly, the set $\mathcal{N}_{\beta,\lambda} = \{ h(\mathbf{z})| \int d\mathbf{z} \: h(\mathbf{z}) \rho_{\text{tot}}(\mathbf{z},\beta,\lambda) =0\} $ consists  of all random fields with vanishing  equilibrium average for fixed $\rho_{\text{tot}}(\mathbf{z},\beta,\lambda)$.
The second line, in which the fluctuating internal energy is expressed as the surplus of the conditional total energy relative to the bare environmental energy  superimposed by a fluctuating contribution $h_u(\mathbf{z},\beta.\lambda)$,  is obtained with the help of Eq.~(\ref{H*Htot}).~\footnote{One possible choice of a random field satisfying $h_u(\mathbf{z},\beta,\lambda) \in \mathcal{N}_{\beta,\lambda}$ is given by $h_u(\mathbf{z},\beta,\lambda) = \alpha(\mathbf{x}) \delta H_{SB}(\mathbf{z})$ assigning an arbitrary fraction $\alpha(\mathbf{x})$ of the fluctuation of the interaction energy $\delta H_{SB}(\mathbf{z}) = H_{SB}(\mathbf{z}) -\langle H_{SB}(\mathbf{z})|\mathbf{x} \rangle$ as an internal energy fluctuation~\cite{Talkner16bPRE}.} It is worth noting that 
with the subtraction of the average bare environment energy 
one may assign a {\it finite} energy to the open system even for large environments in the thermodynamic limit. Without this term, the fluctuating internal energy would depend in a sensitive way on irrelevant details of the environment. Moreover,
without it, the average fluctuating energy of the open system would be given by the internal energy of the {\it total} system rather than by the internal energy difference of the total system and the bare environment as required by Eq.~(\ref{XiXi}).\footnote{\label{fn_seifert}At variance with the previous definition of the fluctuating internal energy in \cite{Seifert16PRL} that corresponds to the choice with a vanishing random field $h_u(\mathbf{z},\beta,\lambda) =0$ in Eq. (\ref{ieH*}), in a recent review of stochastic thermodynamics \cite{Seifert19ARCMP} the fluctuating internal energy is proposed to agree with the conditional total system energy $\langle H_{\text{tot}} |\mathbf{x} \rangle$, i.e.,  without the subtraction of the bare bath energy. Hence, small systems also acquire the typically large, possibly even diverging energy of the environment.}

Once a fluctuating internal energy is assigned to the state of the total system, the momentary energy content $g(\mathbf{z},\beta, \lambda)$ of the reservoir can be identified as the difference of the total energy and the fluctuating internal energy, yielding
\begin{equation}
g(\mathbf{z},\beta, \lambda) =H_{\text{tot}}(\mathbf{z},\lambda) - e(\mathbf{z},\beta,\lambda)\:.
\label{g}
\end{equation}

With this assignment one may {\it define} the heat $q$ exchanged with the environment in the course of  a process in which a system parameter change from $\lambda$ to $\lambda'$ leads the total system to move in phase space from the initial point $\mathbf{z}$ to the final point $\mathbf{z'}$ as
\begin{equation}
q = g(\mathbf{z},\beta, \lambda) - g(\mathbf{z'},\beta, \lambda')\:.  
\end{equation}
With this definition, a positive heat corresponds to an energy taken from the environment. The work $w$, which is performed on the open system in the same realization of the process, is, according to Eq.~(\ref{wHtot}), given by the difference of the total Hamiltonians and hence written as
\begin{equation}
w = H_{\text{tot}}( \mathbf{z'}, \lambda') -H_{\text{tot}}( \mathbf{z}, \lambda)\:.
\label{wH'H}
\end{equation}  
The fluctuating work, heat and internal energy then  satisfy the first-law-like relation
\begin{equation}
\Delta e = q +w\:,
\label{eqw}
\end{equation}
which, however, is of little predictive power because both the fluctuating internal energy change and the heat depend on the difference of  $h_u(\mathbf{z},\beta, \lambda)$ and $h_u(\mathbf{z'},\beta, \lambda')$. These are values of almost arbitrary functions, which are restricted only by having vanishing equilibrium averages. Therefore, {\it only} if both the initial and  final states of a system under the influence of forcing are equilibrium states, according to Eq.~(\ref{ieH*}), the dependence of the average internal energies on the functions $h_u(\mathbf{z},\beta, \lambda(t))$, $t=0,\tau$ disappears and consequently the average of heat can be determined from the difference of the final and initial internal energies and the average work done by the force. More can be done only in the weak coupling limit; see Sec.~\ref{weakcoupling}.

Even though, as mentioned at the start of Sec.~\ref{RMP_3}, the properties of the open system, including its dynamics, are identical for a canonical ensemble of large closed systems, and for a single large  system that weakly couples to a superbath at the required temperature, the present definition of heat is restricted to the former situation because otherwise after a sufficiently large time the heat produced in a cyclic process is finally absorbed by the superbath~\cite{Talkner16bPRE}, particularly note footnote 8 therein.     
\subsection{Fluctuating entropy and free energy}\label{Fefe}
Once a particular fluctuating internal energy is chosen, fluctuating free energies $f(\mathbf{z},\beta,\lambda)$ and entropies $s(\mathbf{z},\beta, \lambda)$ may be assigned under the constraint that their equilibrium averages coincide with the respective potentials $F_S(\beta,\lambda)$ and $S_S(\beta,\lambda)$ of the open system, such that
\begin{align}
\label{ffl}
F_S(\beta,\lambda) &= \int d\mathbf{z}  f(\mathbf{z},\beta, \lambda) \rho_{\text{tot}}(\mathbf{z},\beta,\lambda)\:,\\
S_S(\beta,\lambda) &= \int d\mathbf{z}  s(\mathbf{z},\beta, \lambda) \rho_{\text{tot}}(\mathbf{z},\beta,\lambda)\:,
\label{sfl}
\end{align}
where the pdf $\rho_{\text{tot}}(\mathbf{z},\beta,\lambda )$ is given by Eq. (\ref{rtot}).
To obtain a thermodynamically consistent description of the open system we require the validity of Eqs.~(\ref{U}) and (\ref{S}) between the open system's thermodynamic potentials $U_S$ and $S_S$, respectively, and the corresponding free energy  $F_S$, yielding
\begin{align}
&\int d\mathbf{z}  \rho_{\text{tot}}(\mathbf{z},\beta,\lambda) \Big \{ e(\mathbf{z},\beta,\lambda) - \frac{\partial}{\partial \beta} \beta f(\mathbf{z},\beta,\lambda) \Big .\nonumber \\
&\quad \Big . - \beta f(\mathbf{z},\beta,\lambda) \frac{\partial}{\partial \beta} \ln \rho_{\text{tot}} (\mathbf{z},\beta,\lambda) \Big \} =0\:, \label{ref}\\
&\int d\mathbf{z}  \rho_{\text{tot}}(\mathbf{z},\beta,\lambda) \Big \{ s(\mathbf{z},\beta,\lambda) - k_B \beta^2 \big [\frac{\partial}{\partial \beta} f(\mathbf{z},\beta,\lambda) \big . \Big .\nonumber \\
&\quad \Big . \big . + f(\mathbf{z},\beta,\lambda) \frac{\partial}{\partial \beta} \ln \rho_{\text{tot}} (\mathbf{z},\beta,\lambda) \big ]\Big \} =0\:.
\label{rsf}
\end{align}  
Accordingly, one obtains for the fluctuating thermodynamic potentials the consistency relations
\begin{align}
e(\mathbf{z},\beta,\lambda) &= \frac{\partial}{\partial \beta} \beta f(\mathbf{z},\beta,\lambda) \nonumber\\
& \quad - \beta f(\mathbf{z},\beta,\lambda) \bb { H_{\text{tot}}(\mathbf{z},\lambda) -U_{\text{tot}}(\beta,\lambda ) } \nonumber \\
&\quad + h_e(\mathbf{z},\beta,\lambda)\:,
\label{ef}\\ 
s(\mathbf{z},\beta,\lambda) &=k_B \beta^2 \big [ \frac{\partial}{\partial \beta} f(\mathbf{z},\beta,\lambda) \nonumber\\
& \quad - f(\mathbf{z},\beta,\lambda) \big ( H_{\text{tot}}(\mathbf{z},\lambda) -U_{\text{tot}}(\beta, \lambda) \big ) \big ] \nonumber \\
&\quad + h_s(\mathbf{z},\beta,\lambda)\:,
\label{sf}
\end{align}
where the right-hand sides of Eqs.~(\ref{ef}) and (\ref{sf}) contain arbitrary functions $h_{e/s}(\mathbf{z},\beta,\lambda) \in \mathcal{N}_{\beta,\lambda}$. Apparently, the requirement of thermodynamic consistency is not sufficient to assess these functions  other than by mere definitions.  

Making the assumption that the fluctuating thermodynamic potentials do not explicitly depend on the environmental variables $\mathbf{y}$ one is left with consistency conditions of the same type as Eqs. (\ref{ef}) and (\ref{sf}) in which the full phase-space variable $\mathbf{z}$ is replaced by $\mathbf{x}$. Moreover, in both equations the expression in parentheses has to be modified according to
$\bb {H_{\text{tot}}(\mathbf{z},\lambda) - U_{\text{tot}}(\beta,\lambda) } \to \partial\bb { \beta [H^*(\mathbf{x},\beta,\lambda) -F_S(\beta,\lambda) ] }/\partial \beta$~\cite{Talkner16bPRE}. The unknown functions $ h_{u/e/s}(\mathbf{z},\beta,\lambda) \in \mathcal{N}_{\beta,\lambda}$ of the total phase space must then be replaced by  functions  $h_{u/e/s}(\mathbf{x},\beta,\lambda)$ that have vanishing average with respect to the reduced equilibrium pdf $\rho_S(\mathbf{x},\beta,\lambda) = \int d\mathbf{y} \rho_{\text{tot}}(\mathbf{z},\beta,\lambda)$. 

It might be tempting to choose the functions $h_e(\mathbf{x},\beta,\lambda)$ and $h_s(\mathbf{x},\beta,\lambda)$ in such a way that the fluctuating potentials satisfy the same relations as the respective average quantities do and therefore should be related by (i) $e(\mathbf{x},\beta,\lambda) = \partial \bb {\beta f(\mathbf{x},\beta,\lambda)}/\partial \beta$ and (ii) $s(\mathbf{x},\beta,\lambda) = k_B\beta^2 \partial \bb { f(\mathbf{x},\beta,\lambda) }/\partial \beta$~\cite{Seifert19ARCMP}.
To see whether this assumption is compatible with the required relations for the averages, both sides are averaged with respect to the open system equilibrium pdf $\rho_S(\mathbf{x},\beta,\lambda)$.  From the first equation one obtains  $U_S(\beta,\lambda) = \int d\mathbf{x}  \rho_S(\mathbf{x},\beta,\lambda) \partial\bb { \beta f(\mathbf{x},\beta,\lambda) }/\partial  \beta $. Therefrom, together with the thermodynamic consistency equation (\ref{U}) the condition $\int d \mathbf{x} f(\mathbf{x},\beta,\lambda) \partial \rho(\mathbf{x},\beta,\lambda)/\partial \beta =0$ follows, which, in general, does not hold.  The second equation yields the same condition on the fluctuating free energy. This implies that the thermodynamic consistency is violated in general, by both relations (i) and (ii).\footnote{The thermodynamically consistent fluctuating internal energy and entropy postulated by~\citet{Seifert16PRL} imply a deterministic free energy $f(\mathbf{z},\beta,\lambda) = F_S(\beta,\lambda)$~\cite{Talkner16bPRE}.}     
 \citet{Strasberg17PRE} obtained the same inconsistent relations (i) and (ii) are  obtained using an approximate coarse graining procedure of a master equation.  
\subsection{Fluctuating work and heat in open quantum systems}
According to the detailed discussion in Sec.~\ref{open}, work can be understood as the  difference of the results of two energy measurements of the total system. To obtain an analogous definition of  quantum heat, one needs to know a convenient operator $g$ representing the energy content of the environment. Then the difference of the outcomes of two projective measurements of this operator yields the heat, i.e., the energy lost by the environment. 

For processes during which the system is alternately coupled to and decoupled from, environments,\footnote{\label{transp_eng}Such situations are realized in heat and particle exchange between reservoirs~\cite{Andrieux09NJP,Campisi_Hanggi_Talkner11_aRMP,Jeon17NJP,Campisi10PRL} and also in cyclically performing engines~\cite{Kosloff17Entropy,Zheng16PRE,Ding18PRE}. Possible changes of the energy of the total system due to the time dependence of the coupling and decoupling are typically neglected.} the environmental energy is determined by the Hamiltonian $H_B$ of the bare environment. In other situations with a permanent contact of system and environment one may follow the strategy for classical systems based on a fluctuating internal energy as outlined in Sec.~\ref{Fie}. The quantum analog of a fluctuating internal energy is an internal energy operator $e(\beta,\lambda)$ with the property to yield the internal energy $U_S$ on average in thermal equilibrium, i.e. $U_S(\beta,\lambda) = \Tr_{\text{tot}} \bb { e(\beta,\lambda)\: \rho_{\text{tot}}(\beta,\lambda) } = \Tr_{\text{tot}}\bb { \partial \beta H^*(\beta,\lambda)/\partial \beta  \: \rho_{\text{tot}}(\beta,\lambda) }$. As in the classical case, this requirement leaves considerable ambiguity as to the choice of an internal  energy operator, which can be represented as
\begin{equation}
e(\beta,\lambda) = \frac{\partial}{\partial \beta} \beta H^*(\beta,\lambda) + h_e(\beta,\lambda)\:,
\label{eH*h}
\end{equation}             
where $h_e(\beta,\lambda)$ 
is a Hermitian operator with vanishing equilibrium average 
$\Tr_{\text{tot}} h_e(\beta,\lambda) \rho_{\text{tot}}(\beta,\lambda) =0$.
To any internal energy operator there belongs a corresponding operator $g(\beta,\lambda)$ specifying the energy content of the environment with
\begin{equation}
g(\beta,\lambda)  = H_{\text{tot}} - e(\beta,\lambda)\:.
\label{gHe}
\end{equation}
The heat characterizing a particular process can then be operationally defined in terms of two measurements of this environmental energy operator at the beginning and the end of the respective processes. As in the classical case the environmental energy $g(\beta,\lambda)$ and, consequently, the heat  inherit the ambiguity of the internal energy operator.  

If one is interested in the amount of work and heat that is concurrently supplied to the system and exchanged with the environment in the same process, one faces the problem of having to {\it simultaneously} measure the Hamiltonian of the total system and environmental energy operator $g(\beta,\lambda)$. These operators do not, in general, commute with each other in the presence of an interaction between system and environment. Hence, for systems continuously in contact with their environment a process cannot be characterized by a simultaneous specification of work and heat for the same reason that position and momentum cannot be assigned to a quantum particle at the same time. 
The only exception to this rule is realized for a system weakly coupling with its environment, as discussed in more detail later. The formulation of a first law for systems other than weakly interacting quantum ones therefore seems doubtful to us, contrary to a widespread opposing  opinion~\cite{Nieuwenhuizen02PRE,Alicki79JPA,Seifert16PRL}.      

\subsection{Weak coupling}\label{weakcoupling}
As mentioned in Section \ref{Hmf}, in the weak coupling limit~\cite{vanHove57Phys,Davies76Book} one considers an interaction of vanishingly small strength $\kappa$ between system and environment acting on an increasingly long time-scale such that energy may still flow between system and environment and eventually the small system  equilibrates without a noticeable renormalization of the system's Hamiltonian. Technically speaking, the internal energy operator of the system then agrees with the bare system Hamiltonian $e(\beta,\lambda)=H_S(\lambda)$ and the environmental energy operator with its bare Hamiltonian $g(\beta,\lambda) =H_B$~\cite{Talkner09JSM}. Additional small contributions resulting from the interaction can be neglected in any respect other than for the long time dynamics.  This results in the exceptional situation in which quantum work and quantum heat can be determined for the same process by simultaneously measuring $H_{\text{tot}}(\lambda) = H_s(\lambda) + H_B$ and $g(\beta,\lambda)$ yielding
\begin{align}
w &= E_{m'}(\tau) +\epsilon_{\alpha'} -(E_m(0) +\epsilon_\alpha)\:, \label{wEE}\\
q &= \epsilon_\alpha -\epsilon_{\alpha'}\:,
\label{qepseps}
\end{align}
where $E_m(t)$ denotes an eigenvalue of the system Hamiltonian  $H_S(\lambda(t))$, $\epsilon_\alpha$ denotes the eigenvalue of $H_B$ 
emerging in the first measurement and $\epsilon_{\alpha'}$ denotes the corresponding result of the second measurement.  
The sum of heat and work is given by the difference of eigenvalues of the system Hamiltonian $H_S$, which is consistent with $e(\beta,\lambda) = H_S(\lambda)$.

For a force protocol $\Lambda$ extending over the time span $(0,\tau)$ the joint work and heat pdf $p_\Lambda(w,q)$ becomes~\cite{Talkner09JSM}
\begin{equation}
\begin{split}
p_\Lambda(w,q)& = \sum_{\stackrel{m,m'}{\alpha,\alpha'}} \delta(w-E_{m'} - \epsilon_{\alpha'} +E_m + \epsilon_\alpha)\\
&\quad \times \delta (q-\epsilon_\alpha +\epsilon_{\alpha'}) p_\Lambda(m',\alpha';m,\alpha)\:.
\end{split}
\label{pLwq}
\end{equation}       
Here $p_\Lambda(m',\alpha';m,\alpha)$ specifies the joint probability of finding the total system at the energy $E_m(0)$ and the environment at  $e_\alpha$ immediately before the force protocol starts and at $E_{m'}(\tau)$ and $\epsilon_{\alpha'}$ at the end. It can be written as $p_\Lambda(m',\alpha';m,\alpha) = p_\Lambda(m',\alpha'|m,\alpha)p(m,\alpha)$ in terms of  the initial probability distribution of the total system, $p(m,\alpha) = \Tr \Pi_m(0) Q_\alpha \rho_{\text{tot}}$ and the transition probability
$p_\Lambda(m',\alpha'|m,\alpha) = \Tr_{\text{tot}} \Pi_{m'}(\tau) Q_{\alpha'}U_\Lambda \Pi_m(0) Q_\alpha U^\dagger_\Lambda /\Tr_{\text{tot}} \Pi_n(0)$.   
The projection operators onto the eigenspaces of the Hamiltonians $H_S(\lambda(t))$ and $H_B$ are denoted by $\Pi_n(t)$ and $Q_\alpha$, respectively. The time evolution operator 
\begin{equation}
U_\Lambda = \mathcal{T} e^{- i \int_0^\tau dt H_{\text{tot}}/\hbar} 
\label{Utot}
\end{equation}
is governed by the full Hamiltonian of the total system, including the interaction. 
For short processes of duration $\tau$ with $\kappa^2 \tau \ll 1$ the environmental dynamics is unaffected by the interaction and hence, with $\epsilon_l=\epsilon_k$, the heat typically vanishes. Bath transitions and, accordingly, the heat transfer become important  for long-lasting processes with $\kappa^2 \tau \gtrsim 1$.

The joint work and heat pdf, Eq.~(\ref{pLwq}), describing a process controlled by the force protocol $\Lambda$ and starting from a canonical equilibrium state of the total system at the inverse temperature $\beta$ is linked to the according pdf for the reversed protocol $\bar{\Lambda}$ by a Crooks-type relation~\cite{Talkner09JSM} given as
\begin{equation}
p_\Lambda(w,q) = e^{-\beta(\Delta F_S -w)} p_{\bar{\Lambda}}(-w,-q)\:.
\label{CRwq}
\end{equation}
As an immediate consequence one recovers for the marginal work pdf $p_\Lambda(w) = \int d q p_\Lambda(w,q)$ the Crooks relation [Eq.~\ref{Crops}] and the Jarzynski equality [Eq.~\ref{Jop}] for open systems. In contrast, the marginal heat pdf $p^q_\Lambda(q) = \int dw p_\Lambda(w,q)$ does {\it not} obey a fluctuation relation. We further note that the two fluctuation theorems hold only for the work defined as the energy difference of the total system.  In contrast, the joint pdf $p^{\Delta e,q}_\Lambda(\Delta e,q) = \int dw \delta (\Delta e -w -q) p_\Lambda(w,q)$     
 of the difference of the according internal energy $\Delta e = E_{m'}(\tau) -E_m(0)$ and of the heat  satisfies a Crooks-type relation of the form~\cite{Talkner09JSM}
\begin{equation}
 p^{\Delta e,q}_\Lambda(\Delta e,q) = e^{-\beta(\Delta F - \Delta e +q)} p^{\Delta e,q}_{\bar{\Lambda}}(-\Delta e,-q)\:.
\label{Ceq}
\end{equation}
But because of the presence of the heat in the exponent on the right-hand side, one does not obtain a Jarzynski equality in $\Delta e$, other than for sufficiently short protocols for which the heat vanishes but decoherence may already take place~\cite{Smith18NJP}.

Independent of how strong the interaction between a system and its environment is, the time rate of change of the average bare energy $E_S(t) = \Tr_{SB} H_S(t) \rho_{\text{tot}} = \Tr_S H_S(t) \rho_S(t)$ can always be split into two   contributions according to
\begin{equation}
\dot{E}_S(t) = \Tr_S \frac{\partial H_S(t)}{\partial t} \rho_S(t) +\Tr_S H_S(t) \dot{\rho}_S(t)
\label{dE}
\end{equation}
Eq. (\ref{dE}) can  be considered only as a proper formulation of the first law in the weak coupling limit. Only then does the internal energy of the open system coincide with the thermal average of the bare system Hamiltonian. 
Independent of the coupling strength the integral of the first term on the right-hand side, extended over the time span of the force protocol, yields the average work done on the system according to the TPPEMS provided that the initial state  $\rho_{\text{tot}\:0}$ and the initial  total Hamiltonian $H_{\text{tot}}(\lambda(0))$ commute with each other. In general, the integrated second term, however, has  no definite physical meaning.

Beyond the weak coupling limit the lack of a uniquely defined fluctuating internal energy makes it impossible even to assign an average internal energy to anything other than the according thermal equilibrium state. Because, in general, at the end of a force protocol the system is not in an equilibrium state, it is therefore not possible to assign the change of the average internal energy in  the respective process. Consequently,  an average heat also cannot be specified. 
\section{Outlook and conclusions}\label{RMP_6}
The central notion in the thermodynamics of open systems staying in strong contact with the environment is given by the Hamiltonian of mean force, which is defined in terms of an average of the Boltzmann factor over the thermally distributed environmental degrees of freedom.  It provides at the same time the reduced density matrix of the open system and its thermodynamic equilibrium properties, which, in general, are influenced by the environment in a way that they cannot be inferred from the sole knowledge of the reduced density matrix. Owing to the fact that the resulting partition function $Z_S$ of the open system is given by the ratio of the partition functions of the total system and the bare environment, the existence of $Z_S$ and its independence of irrelevant details of the environment is guaranteed. In particular, remote parts of the environment coupling only weekly  to the system do not affect $Z_S$.  A further important consequence of this particular structure is the finding that the thermodynamic potentials as well as all of their derivatives relating to the open system are determined by differences of the respective quantities referring to the total system and the environment. This guarantees the thermodynamic consistency of the thermodynamic potentials and the  validity of the third law. It also may exhibit unusual properties like negative  entropy and negative specific heat without, however, indicating any instabilities of the respective systems. In the case of  negative entropy it indicates that the interaction between system and environment enforces a state with a higher order than in its absence.

The attempt to represent  the thermodynamic internal energy of an open system as an equilibrium average of a fluctuating internal energy in the case of classical systems, or, for quantum systems, as an internal energy operator,  leads to a tremendous ambiguity in the choice of these fluctuating or operator-valued internal energy expressions. Other fluctuating potentials like fluctuating entropy and fluctuating free energy, as well as the corresponding quantum-mechanical operator-valued expressions, are also affected by these ambiguities.  The interpretation of this inconclusiveness as a kind of gauge freedom~\cite{Jarzynski17PRX} seems rather far-fetched. 
Other than in proper gauge theories there is no obvious advantage to considering  gauge-dependent quantities in the present context.
The fact that a fluctuating thermodynamic potential, on the one hand, plays the role of an observable but, on the other hand, depends on the Gibbs state of the total system appears to be a strange mixture of the two fundamentally distinct categories of states and observables.

Because the specification of heat relies on  the division of the internal energy in work and heat, the notion of heat inherits the ambiguity of the fluctuating internal energy. While the work as a fluctuating quantity can be expressed in an experimentally accessible way for classical systems~\cite{Collin05N}, in quantum systems the TPPEMS of the total energy presents a major experimental challenge~\cite{An14NP}. \footnote{The experimental determination of the {\it average} work in terms of the average power already provides a major challenge as it requires a sufficiently frequent repetition of experimental runs to a variable upper time~\cite{Clos16PRL}.}  For quantum systems the concurrent determination of heat and work is additionally hampered by the fact that it relies on  simultaneous measurements of two energy expressions that do not commute, except for systems weakly coupling to an environment.
In the latter case, for systems that couple weakly to their environment the internal energy can be characterized by the bare system Hamiltonian, and the environmental energy is characterized by its bare bath Hamiltonian.

Before closing we note that in this Colloquium
we do not consider  further relations between thermodynamics and information theory  \cite{Vinjanampathy16CP,Strasberg17PRX} other than those between the thermodynamic entropy of an open system  and several information-theoretic notions  in~\ref{statmechpot}, nor did we discuss the related recent  resource theory approach~\cite{Chitambar19RMP}. In this context we stress  that the frequently made identification of information entropy, typically given by a Shannon or von Neumann entropy, with the thermodynamic entropy must be considered with utmost care as it is by no means guaranteed to be correct~\cite{Alicki19JPA,Norton13Entropy,Hanggi15NP,Hanggi16PTRS}. 
A few further aspects of classical systems outside of equilibrium were considered by~\citet{Talkner16bPRE}.

\begin{acknowledgments}
We thank Michele Campisi, Gert-Ludwig Ingold, Juzar Thingna, Prasanna Venkatesh, Gentaro Watanabe and Juyeon Yi for the numerous discussions and collaborations on the thermodynamics of open systems and fluctuation relations.  Valuable discussions with Massimiliano Esposito, Christopher Jarzynski, Abraham Nitzan, Marti Perarnau-Llobet, Udo Seifert, and Philipp Strasberg are gratefully acknowledged.     
\end{acknowledgments}

\bibliography{PT}
\end{document}